\documentclass{aastex63}
\usepackage{bm}
\usepackage{graphics}
\received{}
\revised{}
\accepted{\today}
\submitjournal{ApJ}

\shorttitle{Raman-scattered \ion{He}{2}}
\shortauthors{Choi et al.}

\begin{document}

\title{Line Formation of Raman-Scattered \ion{He}{2} $\lambda$ 4851 in an Expanding 
Spherical \ion{H}{1} Shell in Young Planetary Nebulae}

\correspondingauthor{Hee-Won Lee}
\email{hwlee@sejong.ac.kr}

\author[0000-0002-9040-672X]{Bo-Eun Choi}
\affil{Department of Physics and Astronomy, Sejong University \\}

\author{Seok-Jun Chang}
\affil{Department of Physics and Astronomy, Sejong University \\}
\affil{Korea Astronomy and Space Science Institute \\}

\author{Ho-Gyu Lee}
\affil{Korea Astronomy and Space Science Institute \\}

\author{Hee-Won Lee}
\affil{Department of Physics and Astronomy, Sejong University \\}



\begin{abstract}
We investigate line formation of Raman-scattered \ion{He}{2} at 4851 \AA~ 
in an expanding neutral spherical shell that surrounds a point-like \ion{He}{2} 
source located at the center. A new grid-based Monte Carlo code is used
to take into consideration the \ion{H}{1} density variation along each photon path.
In the case of a monochromatic \ion{He}{2} emission source,
the resultant line profiles are characterized by an asymmetric double peak structure with a tertiary peak and a significant
red tail that may extend to line centers of \ion{He}{2}$\lambda$4859 and H$\beta$. 
The peak separation corresponds to the expansion velocity, which we consider is in the range $20-40{\rm\ km\ s^{-1}}$ 
in this work. Tertiary red peaks are formed as a result of multiple Rayleigh reflections at the inner surface of a hollow spherical shell 
of \ion{H}{1}. Due to a sharp increase of scattering cross section near resonance, the overall Raman conversion 
efficiency is significantly enhanced as the expansion speed increases.
In the case of a \ion{He}{2} line source with a Gaussian line profile with a full width at half maximum of 
$30 - 70{\rm\ km\ s^{-1}}$, we obtain distorted redward profiles due to increasing 
redward cross section of \ion{H}{1}. A simple application to the young planetary nebula
IC~5117 is consistent with a neutral shell expanding with a speed $\sim 30{\rm\ km\ s^{-1}}$.
\end{abstract}



\section{Introduction} \label{sec:intro}

Stars with mass less than $8\ M_\odot$ lose a significant fraction of their mass via heavy stellar winds in the asymptotic giant branch (AGB) phase, 
before they enter the planetary nebula (PN) stage \citep{hofner18}. 
A PN consists of a hot central star and an expanding shell 
that is photoionized by strong far UV radiation originating from 
the hot central star. High spatial resolution imaging surveys of PNe 
reveal that there appear a variety of nebular morphologies 
including spherical, elliptical, bipolar and point-symmetrical shells \citep{sahai11, hsia14}. 
According to the generalized interaction wind model, the various shell structures may result from the interaction of a fast wind from
the hot central star and a slow wind formed in the AGB phase \citep{kwok78}.
Depending on the column density along a line of sight from the central star of a PN, the surrounding nebula can
be either ionization bounded or matter bounded. In the former case, the distribution and the kinematics of \ion{H}{1} 
in the circumnebular region are very important to understand the mass loss process in the AGB stage and the evolution of PNe.

There have been a number of attempts to detect \ion{H}{1} in PNe using 21 cm hyperfine structure line.
It is a very difficult task to measure the \ion{H}{1} mass in a PN using 21 cm line due to confusion 
arising from the galactic emission.
The first successful detection of atomic hydrogen 
in planetary nebulae was made in NGC~6302 by \cite{rodriguez85}, 
who measured the \ion{H}{1} mass $\sim 0.06 \ M_{\odot}$. 
\cite{rodriguez02} also reported the detection \ion{H}{1} emission
from the Helix Nebula (NGC 7293) noting that the radial velocity of \ion{H}{1} ranges $\sim 50{\rm\ km\ s^{-1}}$. 

Another powerful tool to probe \ion{H}{1} components is Raman-scattered features formed through Raman scattering 
of far UV photons with hydrogen atoms. Nonrelativistic interaction of a far UV photon and an atomic electron 
can be classified into Rayleigh and Raman scattering. In the former case, the initial and final states of the electron are the same so that 
the wavelengths of the incident and outgoing photons are identical. In the case of Raman scattering, 
the final electronic state differs from the initial state resulting
in an outgoing photon with wavelength different from that of the incident photon.
In the vicinity of mass losing stellar objects, Raman scattering is known
to operate, when far UV radiation more energetic than Ly$\alpha$
is incident on a thick neutral region. Raman scattered features therefore
may provide extremely useful information 
on the distribution and kinematics of neutral material associated with
the mass loss processes occurring in the late stage of stellar evolution.

The first astrophysical identification of Raman scattering with atomic hydrogen was proposed in symbiotic stars,
which are wide binary systems of a mass losing giant and a hot star, usually a white dwarf \citep{kenyon86}. 
About a half of symbiotic stars show emission features at 6830 \AA, 
and at 7088 \AA\ with width $\Delta v \sim 100\ \rm km\ \rm s^{-1}$ \citep{allen80}. 
\cite{schmid89} proposed that they are formed as a result of Raman scattering of far UV
doublet \ion{O}{6}~$\lambda\lambda$1032 and 1038 with atomic hydrogen. 
Symbiotic stars are ideal objects for Raman scattering, because a thick neutral region is found
near the mass losing giant and copious \ion{O}{6} line photons are generated around the mass
accreting white dwarf. 

Another interesting application of atomic Raman scattering is found in far UV \ion{He}{2} emission lines. Hydrogen-like ions share
the same energy level structure with level spacing proportional to $Z^2(n_1^{-2}-n_2^{-2})$, where $n_1$ and $n_2$
are principal quantum numbers and $Z$ is the atomic number. In particular, \ion{He}{2} emission lines resulting 
from transitions to $n=2$ levels from $n=2k$ levels have wavelength slightly shortward of \ion{H}{1} Lyman series 
associated with the $kp$ state. Therefore, Raman-scattered \ion{He}{2} features are formed blueward of \ion{H}{1} 
Balmer lines. For example, far UV \ion{He}{2} emission lines at 1025 \AA, 972 \AA\ and 949 \AA\ 
are Raman-scattered to form emission features at 6540 \AA, 4851 \AA\ and 4332 \AA, respectively.

The first detection of Raman-scattered \ion{He}{2} feature was reported by \cite{vangroningen93}, who found a broad emission 
feature at 4851 \AA\ in the symbiotic nova RR~Telescopii. Subsequently Raman-scattered \ion{He}{2} features were found
in the symbiotic stars V1016~Cygni, HM~Sagittae, and V835~Centauri \citep{birriel04, lee01}. Assuming the case B recombination theory
is valid for \ion{He}{2} lines, one can deduce the flux of \ion{He}{2}$\lambda$972 by measuring the flux of 
optical \ion{He}{2}$\lambda$4859 or \ion{He}{2}$\lambda$4686 and hence calculate the Raman conversion efficiency 
defined as the photon number ratio of \ion{He}{2}$\lambda$972 and Raman-scattered \ion{He}{2}$\lambda$4851.

Young planetary nebulae also satisfy the requirement for operation of Raman scattering of \ion{He}{2} with the presence of 
a strong \ion{He}{2} emission region near the hot central white dwarf surrounded by a thick neutral region as a result 
of recent mass loss.
\cite{pequignot97} reported their discovery of Raman-scattered \ion{He}{2} in the young planetary nebula NGC~7027.
Raman \ion{He}{2} features were also found in a number of young planetary nebulae 
including NGC~6302 \citep{groves02}, IC~5117 \citep{lee06} and NGC~6790 \citep{kang09}.
It is quite notable that molecular lines have been detected 
in all these young planetary nebulae with Raman-scattered \ion{He}{2}. 
For example, $\rm H_2$ was observed in NGC~7027, NGC~6302, and IC~5117 \citep{kastner96}, 
and polycyclic aromatic hydrocarbon (PAH) was detected in NGC~6790 \citep{smith08}. The presence of molecular components 
in these objects indicates their recent entrance of the planetary nebula stage. 
In this regard, Raman-scattered \ion{He}{2} features are great implement to study the circumnebular \ion{H}{1} region of young PNe.


\cite{jung04} performed Monte Carlo simulations to investigate line formation of Raman-scattered \ion{He}{2} in a static 
neutral medium illuminated by a \ion{He}{2} emission source. Basic results obtained from studies of line
formation in static media have been used to derive the content of neutral material surrounding the hot central stars
of planetary nebulae.
However, in young planetary nebulae the neutral and molecular components are supposed to move radially outward
with a speed that is comparable to the escape velocity $v_{\rm esc}\sim 20{\rm\ km\ s^{-1}}$ of a typical giant star. 
In this environment, 
the redward line center shift
will be enhanced leading to significant distortion in the line profile.
In this article, we use a grid-based Monte Carlo code to investigate line formation of Raman-scattered \ion{He}{2}$\lambda$4851 
with moving \ion{H}{1} spherical shell. 

This paper is organized as follows. 
Basic atomic physics, scattering geometry and Monte Carlo procedures
are described in Section \ref{sec:atomic}. 
We present our results in Section~\ref{sec:results}, which is followed
by summary and discussion in the final section.

\section{Atomic Physics and Radiative Transfer} \label{sec:atomic}

\subsection{Cross Section and Branching Ratios} \label{subsec:cross}

Regarded as a two-body system, the reduced mass of \ion{He}{2} is slightly heavier than that of \ion{H}{1} by a factor 
$3m_e/(4m_p) \sim 4.08 \times 10^{-4}$, where $m_e$ and $m_p$ are the electron and proton masses, respectively.
This slight excess in reduced mass leads to the energy level spacing 
of \ion{He}{2} larger than that of \ion{H}{1} by a factor slightly in excess of 4 \citep{lee01}. 
More quantitatively, the line center of \ion{He}{2}$\lambda$972
is blueward of \ion{H}{1} Ly$\gamma$ by $\Delta V=-124{\rm\ km\ s^{-1}}$.

The cross sections for Rayleigh and Raman scatter are given as an infinite sum of the probability amplitudes
associated with all bound and free $p$ states, which is known as the Kramers-Heisenberg formula \citep[e.g.][]{bethe67, sakurai67}
\begin{eqnarray}
{d\sigma \over d\Omega} &=&r_e^2 
\left({\omega' \over \omega}\right)\Bigg| \delta_{AB}({\bm\epsilon^{(\alpha)}}\cdot {\bm\epsilon^{(\alpha')}})+
{1\over m_e} \sum_I \left(
{({\bf p}\cdot {\bm \epsilon^{(\alpha')}})_{BI}
({\bf p}\cdot {\bm \epsilon^{(\alpha)}})_{IA}
\over E_I -E_A-\hbar\omega}
\right.
\nonumber \\
&+& \left.{({\bf p}\cdot{\bm\epsilon^{(\alpha)}})_{BI}
({\bf p}\cdot{\bm\epsilon^{(\alpha')}})_{IA}
\over E_I -E_A+\hbar\omega'}
\right) \Bigg|^2.
\end{eqnarray}
Here, $r_e= e^2 / m_e c^2 = 2.82 \times 10^{-13} \rm\ cm$ is the classical electron radius. 
The initial and final electronic states are represented by $A$ and $B$, and $I$
stands for an intermediate state, which is one of $p$ states in the case
of electric dipole processes.
In the case of 
Rayleigh scattering, $A=B=|1s>$, and in the case of
Raman scattering, $B$ is one of $2s, 3s$ and $3d$. 
The Kronecker delta symbol $\delta_{AB}$ vanishes in the case of Raman scattering, 
whereas it becomes unity in the case of Rayleigh scattering.
The angular frequencies of the incident and outgoing photons 
are denoted by $\omega$ and $\omega'$, respectively. 
The polarization vectors of the incident and outgoing photons are 
denoted by ${\bm\epsilon^{(\alpha)}}$ and ${\bm\epsilon^{(\alpha')}}$, 
respectively.

The term $({\bf p}\cdot {\bm \epsilon^{(\alpha)}})_{IA}$ is the matrix element 
of the momentum operator component along the polarization 
vector ${\bm\epsilon^{(\alpha)}}$ between the states $I$ and $A$, which is 
proportional to the dipole transition amplitude.
The explicit values of the matrix element can be found in \cite{karzas61}.
These matrix elements can also be computed using the Dalgarno-Lewis method where the cross sections are expressed 
as a convolution
of the final state wavefunction and the Green's functions associated with the Rayleigh and Raman interactions
\citep[e.g.][]{sadeghpour92}.

\begin{figure}
\epsscale{0.7}
\plotone{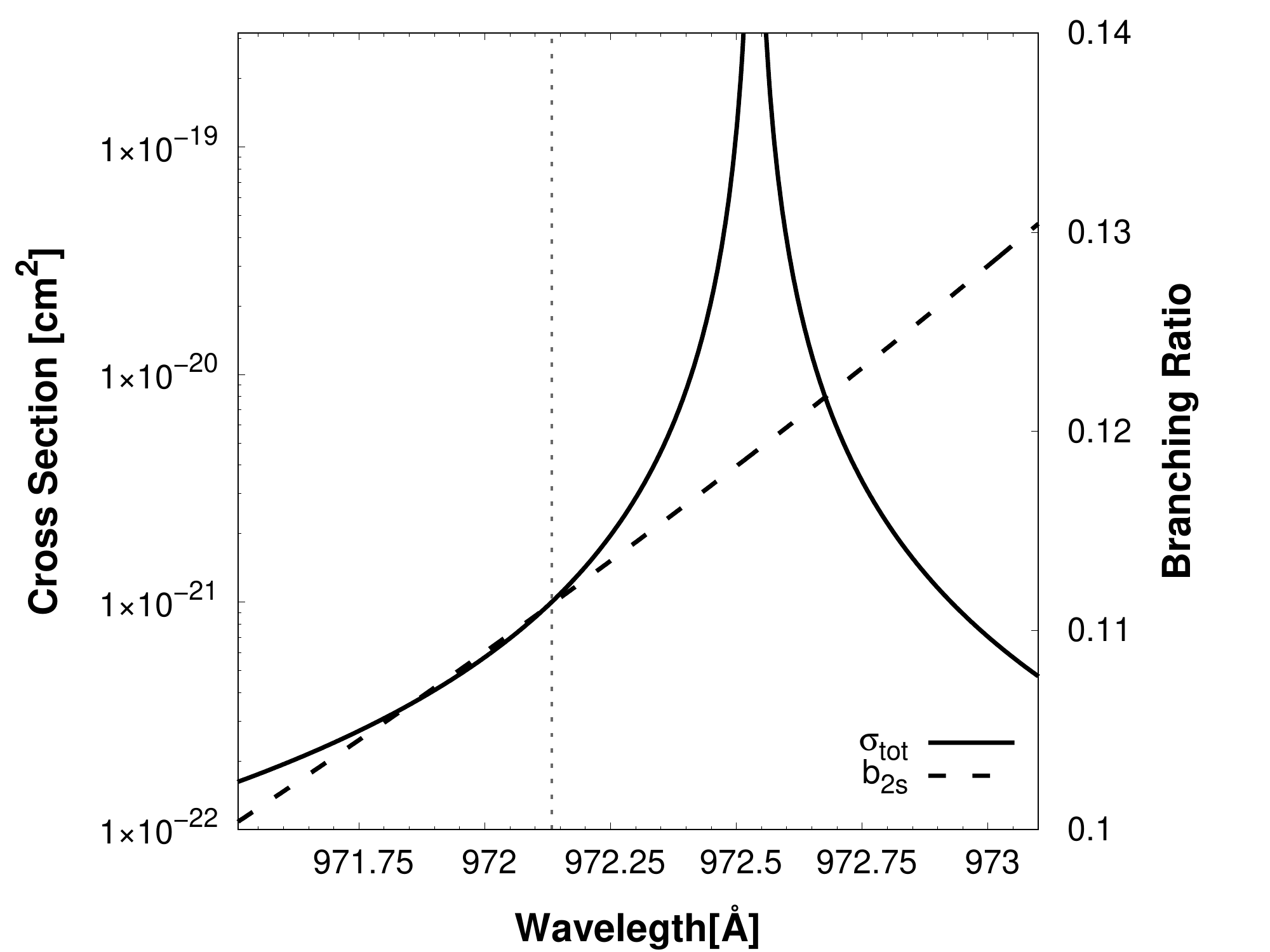}
\caption{Cross section and branching ratio of far UV radiation near
\ion{He}{2}$\lambda$972. The solid curve shows the total cross section 
and dashed line shows the branching ratio of transition to the $2s$ level. 
The dotted vertical line shows the rest wavelength of \ion{He}{2}$\lambda$972 
emission line. 
} 
\label{fig:cross}
\end{figure}

In Fig.~\ref{fig:cross}, we show the total cross section $\sigma_{\rm tot}$ , where the total cross section is the sum of cross sections 
of Rayleigh and Raman scattering. In the same figure, the branching ratio $b_{2s}$ for Raman scattering into the $2s$ state is also shown by a dotted line. 
In the wavelength range
shown in the figure, the branching ratio into the $2s$ state ranges between 0.1 and 0.13. It is clearly notable that both the cross section and the branching 
ratio are rapidly increasing in the wavelength range shown in the figure.

For Raman-scattered \ion{He}{2}$\lambda$4851, proximity to $1s-4p$ resonance of \ion{H}{1} implies that
the dominant contribution comes from the probability amplitude associated with the dipole expectation values 
corresponding to the $1s\to 4p\to 2s$ transition.
If we approximate the total scattering cross section (sum of Rayleigh and Raman scattering cross sections) by a Lorentzian function 
neglecting $p$ states other than $4p$, we obtain
\begin{equation}
\sigma_{\rm Ram}(\lambda) \simeq \sigma_0 \left({\lambda_0\over \lambda-\lambda_0}\right)^2,
\end{equation}
where $\sigma_0=0.17\times 10^{-27}{\rm cm^2}$ \citep{jung04}. At the line center of \ion{He}{2}$\lambda$972, it turns out that
the total scattering cross section is
\begin{equation}
\sigma_{\rm tot, HeII972} = 9.1\times 10^{-22}{\rm\ cm^2},
\end{equation}
with the branching ratio into the final $2s$ state being 0.11.
Therefore, we may roughly expect that Raman scattering may result once out of 9 Rayleigh scattering events.

From the energy conservation, the wavelength $\lambda_f$ of a Raman-scattered photon is related to that $\lambda_i$
of the incident far UV photon by
\begin{equation}
{1\over \lambda_f} = {1\over \lambda_i} - {1\over \lambda_{\rm Ly\alpha}},
\label{eq:freq}
\end{equation}
where $\lambda_{\rm Ly\alpha}$ is the line center wavelength of \ion{H}{1} Ly$\alpha$. In the case of \ion{He}{2}$\lambda$972, the
line center wavelength averaged over all fine structures is $\lambda_{i,\rm HeII972}=972.134{\rm\ \AA}$ \citep{lee06}, resulting in the atomic
line center of Raman-scattered \ion{He}{2}$\lambda$4851 at 
\begin{equation}
\lambda_{\rm c,4851}= 4851.29 {\rm\ \AA}
\label{eq:atomic_cen}
\end{equation}
with a consideration of the air index of refraction
$n_{\rm air}=1.000279348$ and $\lambda_{\rm Ly\alpha}=1215.67 $ \AA\ \citep{jung04}. From the line center of H$\beta$,
the Doppler factor of the Raman-scattered \ion{He}{2}$\lambda$4851 is
\begin{equation}
\Delta V_{\rm Ram4851} \simeq -619 {\rm\ km\ s^{-1}},
\end{equation}

One important spectroscopic property of Raman-scattered features lies with line broadening by a factor $\lambda_o/\lambda_i$,
which is attributed to the inelasticity of scattering. Noting that $\lambda_{\rm Ly\alpha}$ is a constant, the line broadening relation for Raman scattering
is obtained by differentiating Eq.~\ref{eq:freq}, which leads to
\begin{equation}
{\Delta \lambda _f \over \lambda _f} =\left( {\lambda _f \over \lambda _i}\right)~
{\Delta \lambda _i \over \lambda _i}.
\label{eq:broadening}
\end{equation}

For example, if a hydrogen atom moves away from a \ion{He}{2} ion with a speed $\Delta v=+20{\rm\ km\ s^{-1}}$, the wavelength
shift $\Delta\lambda_f$ of a Raman-scattered \ion{He}{2}$\lambda$4851 is
\begin{equation}
\Delta\lambda_f = 1.62 {\rm\ \AA}
\end{equation}
which would correspond to a new Doppler shift $\Delta v=+100{\rm\ km\ s^{-1}}$ with respect to the Raman-scattered \ion{He}{2} line center.
If the relative speed is
$\Delta v=+124{\rm\ km\ s^{-1}}$ between \ion{H}{1} and \ion{He}{2}, then \ion{He}{2}$\lambda$972 photon is regarded
as a Ly$\gamma$ photon in the rest frame of the hydrogen atom, and Raman-scattering becomes effectively resonance scattering of Ly$\gamma$. 

A far UV \ion{He}{2}$\lambda$972 line photon incident on a thick neutral region, it may be Rayleigh-scattered several times before it may
escape from the region. Rayleigh scattering of a far UV photon is described as elastic in the rest frame of the scatterer, so that
in the observer's rest frame the wavelength of Rayleigh-scattered photon changes according to
\begin{equation}
\Delta\lambda =\lambda_i \left[ ( {\bf\hat k}_i - {\bf\hat k}_f )\cdot {\bf v_{\rm HI}} \over c \right],
\end{equation}
where ${\bf\hat k}_i$ and ${\bf\hat k}_f$ are unit wavevectors associated with the incident and outgoing Rayleigh-scattered photons,
${\bf v}_{\rm HI}$ is the velocity of the hydrogen atom 
in the observer's rest frame. From this relation, a Rayleigh-reflected
photon from a hydrogen atom moving with a speed $v_{\rm HI}=40{\rm\ km\ s^{-1}}$ will acquire a Doppler factor $80{\rm\ km\ s^{-1}}$
and subsequent Raman-scattering will result in a red shifted optical photon with $\Delta V\sim 400{\rm\ km\ s^{-1}}$
from the atomic line center of Raman-scattered \ion{He}{2}. Several Rayleigh-reflections are sufficient to achieve wavelength shift
to reach \ion{H}{1} Ly$\gamma$ resonance to emit a H$\beta$ line photon as a result of final Raman scattering. In this particular case,
Raman conversion efficiency becomes significantly enhanced because of enormous cross section at resonance.

\subsection{Scattering Geometry} \label{sec:model}

\begin{figure}
\centering
\includegraphics[scale=0.5]{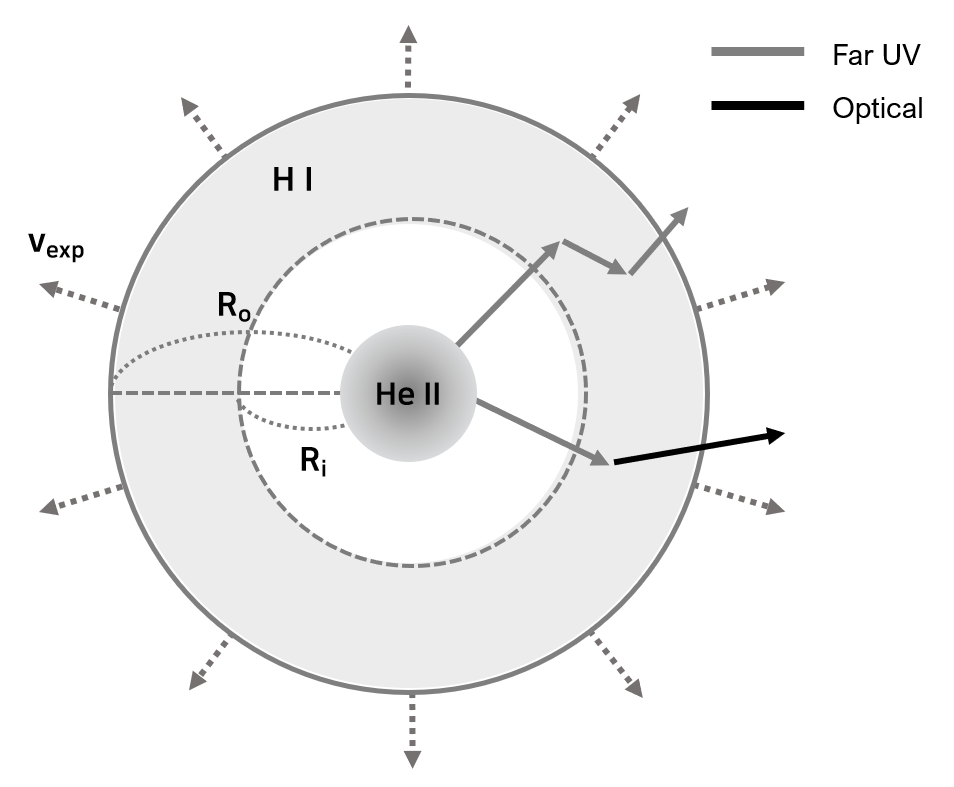}
\caption{
Schematic illustration of the simulation geometry.
The \ion{H}{1} region is assumed to be a spherical 
shell with inner and outer radii of $R_{\rm i}$ and $R_{\rm o}$, respectively. 
A point-like \ion{He}{2} emission source is located at the center.
The scattering region expands radially
from the \ion{He}{2} source with a constant velocity $v_{\rm exp}$.
\ion{He}{2} photons may escape through either Rayleigh or Raman scattering. 
The black arrow shows an escaping optical Raman photon, whereas
the gray arrows represent paths traversed by Rayleigh scattered far UV photons.}
\label{fig:scat_geo}
\end{figure}

Near IR emission line at 2 $\mu{\rm m}$ of ${\rm H}_2$ was found in the planetary nebula NGC~7027 by \cite{treffers76}.
Detection of molecular emission from planetary nebulae strongly implies that the circumnebular region is ionization-bounded
rather than matter-bounded \citep{dinerstein91}.
Circumnebular neutral hydrogen \ion{H}{1} has been detected in several young planetary nebulae through
21 cm radio observation. \cite{taylor89} carried out 21 cm observations of the planetary nebula IC~418 using the VLA to detect an 
absorption feature due to the expanding nebula and an emission consistent with a total \ion{H}{1} mass $\sim 0.07 {\rm\ M_\odot}$
assuming a spin excitation temperature of $\sim 10^3{\rm\ K}$.

Fig.~\ref{fig:scat_geo} is a schematic illustration of the scattering geometry considered in this work. The \ion{He}{2} emission
source is located at the center surrounded by a neutral spherical shell with inner and outer radii $R_{\rm i}$ and $R_{\rm o}$,
respectively. We assume that the spherical shell moves radially outward with a single speed $v_{\rm exp}$ in such a way
that the mass flux is conserved at each radial coordinate $r$. 
That is, the \ion{H}{1} number density $n_{\rm HI}(r)$ is given by
\begin{eqnarray}
n_{\rm HI}(r) &=& {\dot{M} \over {4 \pi r^2 \mu_{\rm a} m_{\rm p} v_{\rm exp}}}
\nonumber \\
&=& 1.34 \times 10^5\ {\rm cm^{-3}}\
\left({\dot{M} \over {10^{-5}\ \rm M_\odot\ yr^{-1}}}\right)
\left({r \over {10^3 \ {\rm au}}}\right)^{-2} 
\left({v_{\rm exp} \over 10\ {\rm km\ s^{-1}}}\right)^{-1}
\label{eq:density}
\end{eqnarray}
where $\dot M$ is the mass loss rate and $\mu_{\rm a}$ is the mean molecular weight.
For simplicity, we set $\mu_{\rm a} = 1$ assuming that the scattering region purely consists of atomic hydrogen.

We measure the \ion{H}{1} column density $N_{\rm HI}$ of the \ion{H}{1} shell along the radial direction so that 
\begin{equation}
N_{\rm HI}=\int_{R_i}^{R_o} n_{\rm HI}(r) drN_{\rm HI,0}\left( 1-{R_i\over R_o} \right).
\end{equation}
Here, the characteristic \ion{H}{1} column density can be expressed as
\begin{equation}
N_{\rm HI,0}=2\times 10^{21} \dot M_{-5}\ r_{-3}^{-2}\ v_{\rm exp,10}^{-1} {\rm\ cm^{-2}},
\end{equation}
in terms of dimensionless quantities $\dot M_{-5}=\dot M/(10^{-5}{\rm\ M_\odot\ yr^{-1}})$,
$r_{-3}=r/(10^3{\rm\ au})$ and $v_{\rm exp,10}=v_{\rm exp}/(10{\rm\ km\ s^{-1}})$. 
The \ion{H}{1} mass of the neutral shell is given by
\begin{equation}
M_{\rm HI}= (5\times 10^{-3}{\rm\ M_\odot})\ \dot M_{-5}\ v_{\rm exp,10}^{-1} 
\left({R_o\over 10^3 {\rm\ au}}\right)
\left( 1-{R_i\over R_o} \right).
\end{equation} 

In this work, we assume that the \ion{He}{2} emission source is unpolarized 
and isotropic to focus on basic physics of line profile formation. 
We consider two cases, where the \ion{He}{2} emission source is 
monochromatic at the line center in the first case and in the second case
the emission line profile is described by a Gaussian function.

\subsection{Grid-based Monte Carlo Approach} \label{sec:method}

In order to describe Rayleigh and Raman scattering in an expanding neutral medium, a new grid-based Monte Carlo code 
was developed, where a Cartesian coordinate system is adopted and the scattering region is divided into a large number 
of cubes of equal size. 
The \ion{He}{2} line source is located at the center of the coordinate 
system and uniform physical properties such as \ion{H}{1} number density 
$n_{I}$ and velocity ${\bf v}_{I}$ are assigned to each cube~$I$ 
with the center coordinate $(x_i, y_j, z_k)$.

For cube~$I$ with radial distance $r=(x_i^2+y_j^2+z_k^2)^{1/2}$, the 
\ion{H}{1} number density
$n_{I}$ is given by Eq.~(\ref{eq:density}) if $R_{\rm i}<r<R_{\rm o}$. 
Otherwise, we set $n_{I}=0$.
The expanding velocity of cube~$I$ is also given by
\begin{equation}
{\bf v}_{I} = v_{\rm exp}{x_i {\bf\hat x}+ y_j {\bf\hat y}+z_k {\bf\hat z} \over r},
\end{equation}
in accordance with our assumption of spherical inertial expansion. 

Now we consider a \ion{He}{2}$\lambda$972 photon propagating from a starting point ${\bf r}_p$ 
with a unit wavevector ${\bf\hat k}$. We determine all the non-empty cubes that lie along the semi-infinite ray 
starting at ${\bf r}_p$ in the direction ${\bf\hat k}$. Labeling these cubes with subscript $I$, we compute 
the scattering optical
depth $\tau_I$ corresponding to the photon path inside cube~$I$ between the entering and exiting points.
In the rest frame of cube~$I$, the \ion{He}{2} photon is treated as monochromatic with frequency
\begin{equation}
\omega_I = \omega_0 (1 + {\bf v}_I\cdot {\bf\hat k}/c),
\end{equation}
where ${\bf\hat k}$ is the unit wavevector of the photon. Using this frequency, we obtain an appropriate cross
section $\sigma_{\rm tot}(\omega_I)$.
With this cross section, the scattering optical depth $\tau_I$ is given by 
\begin{equation}
\tau_I = n_{I} \sigma_{\rm tot}(\omega_I) l_I
\end{equation}
where $l_I$ is the path length through the cube. By summing all $\tau_I$, we obtain $\tau_{\infty}$, the scattering optical depth
to an observer at infinity.

A line photon is supposed to traverse a physical distance $l$ that corresponds to an optical depth $\tau$
given by
\begin{equation}
\tau = -\ln r_{\rm p}
\end{equation}
where $r_{\rm p}$ is a uniform random deviate in the range between 0 and 1. If $\tau>\tau_\infty$, then the line photon
passes through the neutral region and reaches the observer as a far UV \ion{He}{2}$\lambda$972 photon.
Otherwise, the photon is scattered at some cube $J$, where the cumulative optical depth $\tau_I$ just exceeds $\tau$.
Linearly interpolating the optical depths at the entering and exiting points of cube $J$, we determine 
the scattering site inside cube $J$.

In this work, the geometry is spherically symmetric so that no consideration is given to polarization. According to \cite{stenflo80}, 
the probabilistic angular distribution of the scattered radiation for Rayleigh and Raman scattering with atomic hydrogen sufficiently 
far from resonance is identical with that for Thomson scattering, which is given as
\begin{equation}
f(\beta)\propto (1+\beta^2).
\end{equation}
Here, $\bm \beta$ is the cosine of the angle between the incident 
and scattered photons.

At the scattering site, the wavevector is chosen according to the scattering phase function and the scattering type is also
determined. In accordance with the branching ratios $b_{r1}, b_{r2}$ and $b_{r3}=1-b_{r1}-b_{r2}$ with $b_{ri}$ branching
ratios for scattering into $n=1$, $n=2$ and $n=3$ levels, respectively.

We regard the neutral shell is optically thin to optical photons, so that 
Raman-scattered photons escape from the region without any further interaction. 
If the scattering is Rayleigh, the procedure is repeated until 
the far UV \ion{He}{2} photon escapes either as a Raman-scattered 
optical photon or as a Rayleigh-scattered \ion{He}{2} photon 
when $\tau>\tau_\infty$.

\section{Results} \label{sec:results}
\subsection{Monochromatic source} \label{sec:monochro}

\begin{figure}
\plottwo{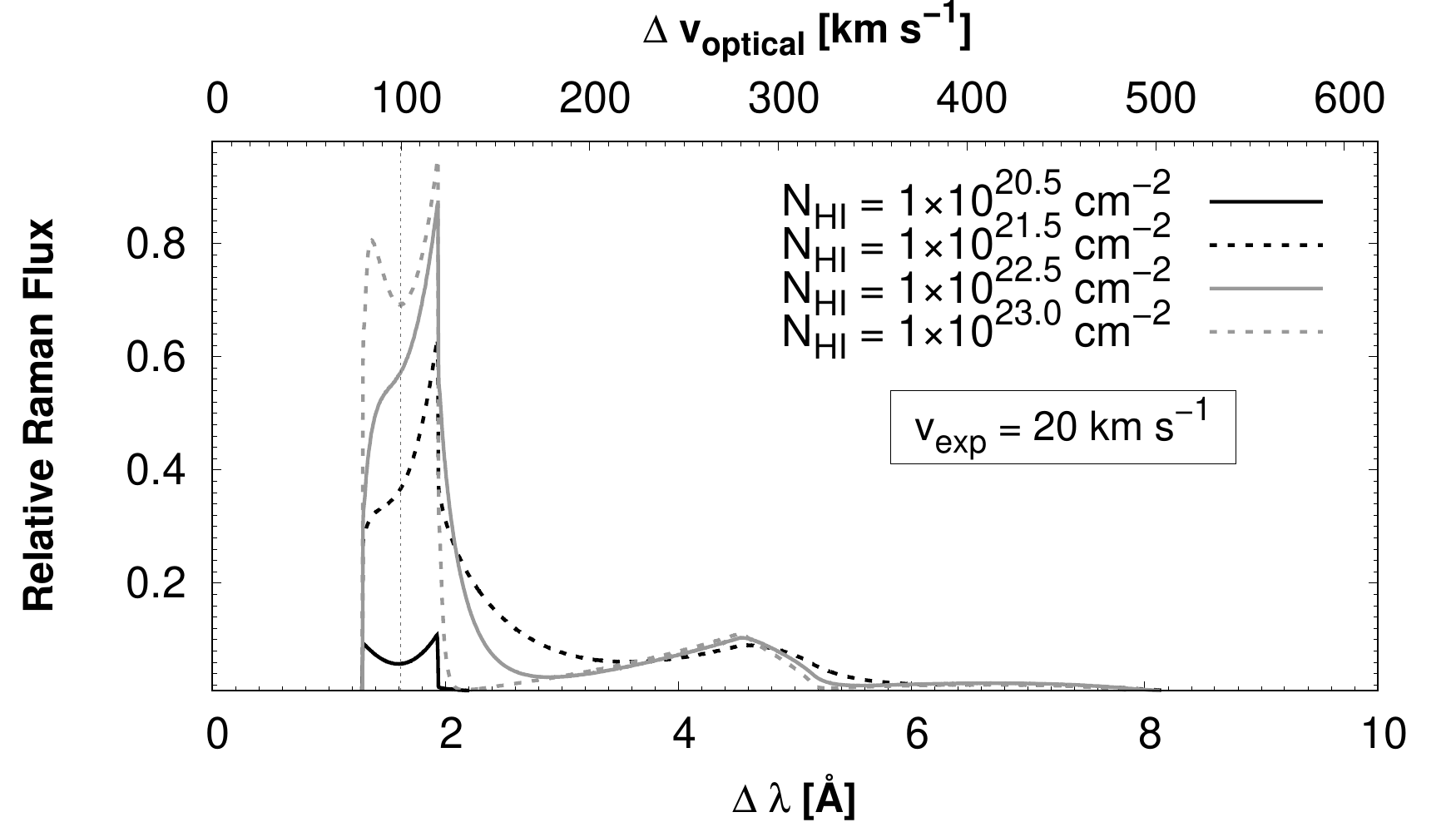}{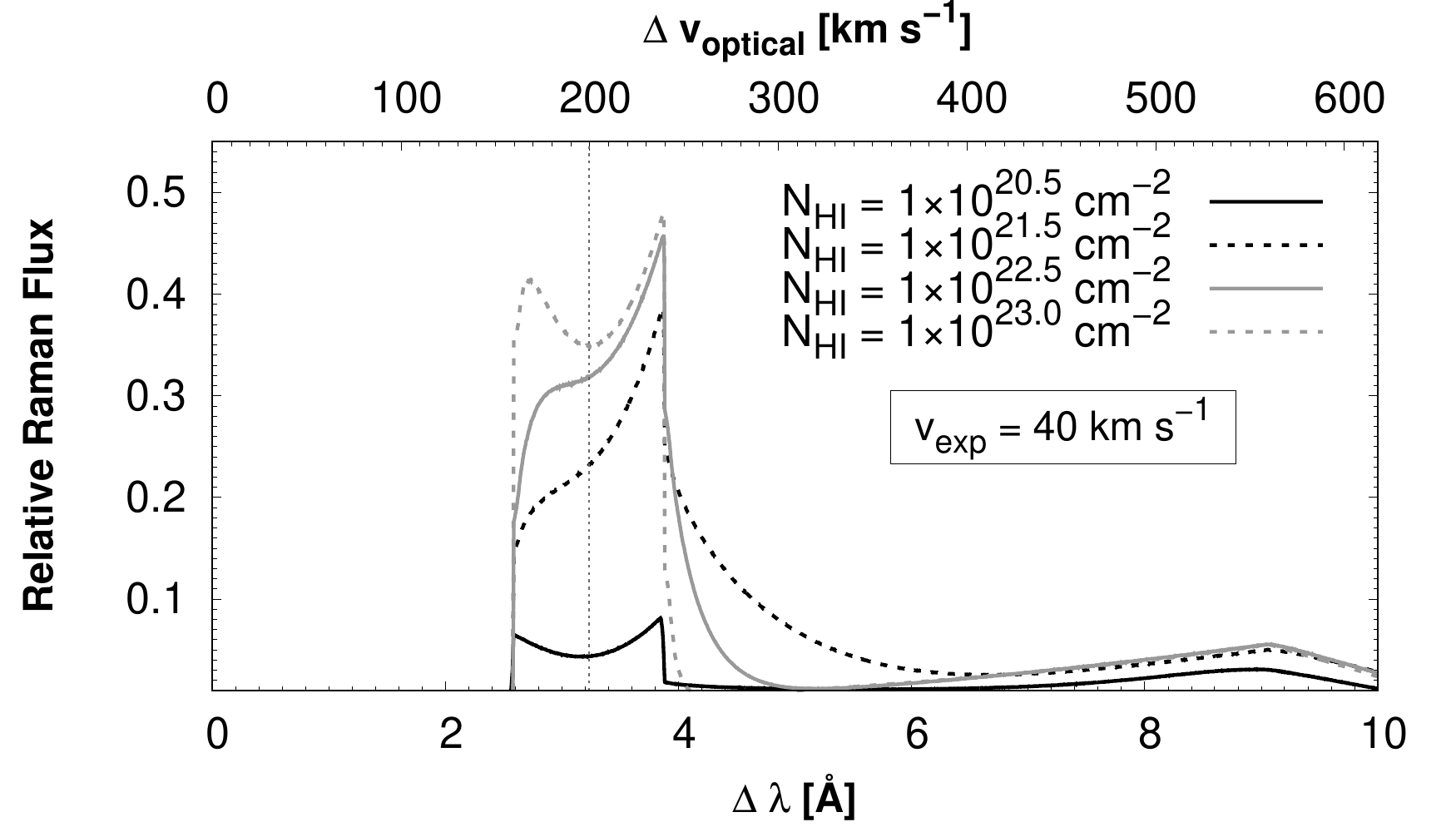}
\caption{
Line profiles of Raman \ion{He}{2} $\lambda$4851 obtained from our Monte Carlo simulations
for various column densities with two values $v_{\rm exp} = 20{\rm\ km\ s^{-1}}$ (left panel) and $40{\rm\ km \ s^{-1}}$ (right panel) 
of the expansion speed of the spherical
neutral shell. The horizontal axis represents the wavelength shift from Raman \ion{He}{2} line center
in units of \AA\ (lower axis) and ${\rm km\ s^{-1}}$ with respect to optical 
Raman-scattered line center(upper axis). 
The vertical dotted line in each panel marks the wavelength shift 
that corresponds to $v_{\rm exp}$.
}
\label{fig:mono_column}
\end{figure}

In Fig~\ref{fig:mono_column}, we show the resultant profiles of Raman-scattered \ion{He}{2}$\lambda$4851 formed in an expanding 
spherical \ion{H}{1} shell surrounding a monochromatic point-like \ion{He}{2} emission source located at the center of the shell.
We consider various values $N_{\rm HI}$ of \ion{H}{1} column density ranging from $10^{20.5}{\rm\ cm^{-2}}$ to $10^{23}{\rm\ cm^{-2}}$.
The left and right panels of Fig~\ref{fig:mono_column} correspond to the cases of the expansion velocity $v_{\rm exp}=
20{\rm\ km\ s^{-1}}$ and $40{\rm\ km\ s^{-1}}$, respectively.

The lower horizontal axis shows the wavelength shift $\Delta\lambda$ from the atomic line center $\lambda_{\rm c,4851}$
of Raman \ion{He}{2}$\lambda$4851 in units of \AA. The upper horizontal axis shows apparent Doppler factor 
$\Delta V$ in units of ${\rm\ km\ s^{-1}}$, which is related to the wavelength shift $\Delta\lambda$ by
\begin{equation}
{\Delta V \over c} = \left( {\lambda_{\rm O} \over\lambda_{\rm I}} \right) 
{\Delta\lambda \over \lambda_{\rm O}}. 
\end{equation}
It should be noticed that the apparent Doppler factor $\Delta V$ approximately correspond to the relative speed of \ion{H}{1} and \ion{He}{2}
multiplied by a factor 5. The vertical dotted line in each panel marks 
the center of the Raman feature located at 
$\Delta V_{\rm c}=(\lambda_{\rm O}/\lambda_{\rm I})
v_{\rm exp}$, which is $\sim 100{\rm\ km\ s^{-1}}$
and $200{\rm\ km\ s^{-1}}$ for $v_{\rm exp}=20$ and $40{\rm\ km\ s^{-1}}$.
In this article, we call the center wavelength $\lambda_{\rm c}$ corresponding to $\Delta V_{\rm c}$ the `Raman line center.'

A spherical shell with $N_{\rm HI}=10^{20.5}{\rm\ cm^{-2}}$ expanding with $v_{\rm exp}<100{\rm\ km\ s^{-1}}$ is optically thin for both
Rayleigh and Raman scattering, and therefore resultant profiles are double-peaked consisting of two peaks of similar strength. It is also notable 
that there appears a faint extended red tail. 
In this case, a significant fraction of incident far UV photons penetrate the \ion{H}{1} region with no interaction or escape the scattering region 
after a single Rayleigh scattering, leading to formation of a very weak Raman feature. 
The Raman feature is contributed dominantly by singly Rayleigh scattered radiation. 
Since the phase function for Rayleigh and Raman scattering sufficiently far from resonance is symmetric with respect to the incidence direction, 
forward and backward scatterings are equally probable, resulting in an almost symmetric double peak structure with 
blue and red peaks located at 
$\Delta V_{\rm b}=\Delta V_{\rm c} - v_{\rm exp}$, and $\Delta V_{\rm r}=\Delta V_{\rm c} + v_{\rm exp}$, respectively

As $N_{\rm HI}$ is increased to $N_{\rm HI}\sim 10^{22}{\rm\ cm^{-2}}$, the blue peak becomes significantly weaker relative to the red counterpart accompanied 
by the enhancement of the red extended tail. When a \ion{He}{2}$\lambda$972 line photon is Rayleigh scattered in an expanding \ion{H}{1} region, it is redshifted 
in the rest frame of the scattering hydrogen atom. Because Rayleigh scattering cross section increases sharply as a function of wavelength, 
the expanding neutral medium becomes moderately optically thick with respect to Rayleigh scattering. Several Rayleigh scatterings before a final Raman scattering 
provide sufficient redward push in wavelength space leading to frequency redistribution from the blue peak to the extended red tail.

However, in the case at very high \ion{H}{1} column density $N_{\rm HI} \ge 10^{22.5} \rm cm^{-2}$, 
the scattering region is optically thick to most \ion{He}{2}$\lambda$972 line photons for which Rayleigh scattering is effectively local due to
short mean free path. Therefore, redward diffusion in wavelength space 
is severely restricted 
because of the small velocity difference
in a localized region. Consequently, considerably symmetric double peak 
profiles are restored in a neutral region with very high $N_{\rm HI}$, 
which is shown in Fig~\ref{fig:mono_column}.

A similar behavior is also found in the profiles shown in the right panel 
of Fig~\ref{fig:mono_column}, for which $v_{\rm exp}=40{\rm\ km\ s^{-1}}$. 
Because $v_{\rm exp}$ is twice that for the left panel, 
the Doppler factor $\Delta V_c$ for the 'Raman line center' and the width 
of the main double peak part are twice the counterparts shown in the left panel. 
In addition, as the scattering cross section is higher for large $\lambda$, 
the 
total Raman flux integrated over the entire wavelength interval
is stronger than the counterpart 
in the left panel with the same $N_{\rm\ HI}$. 

\begin{figure}
\plotone{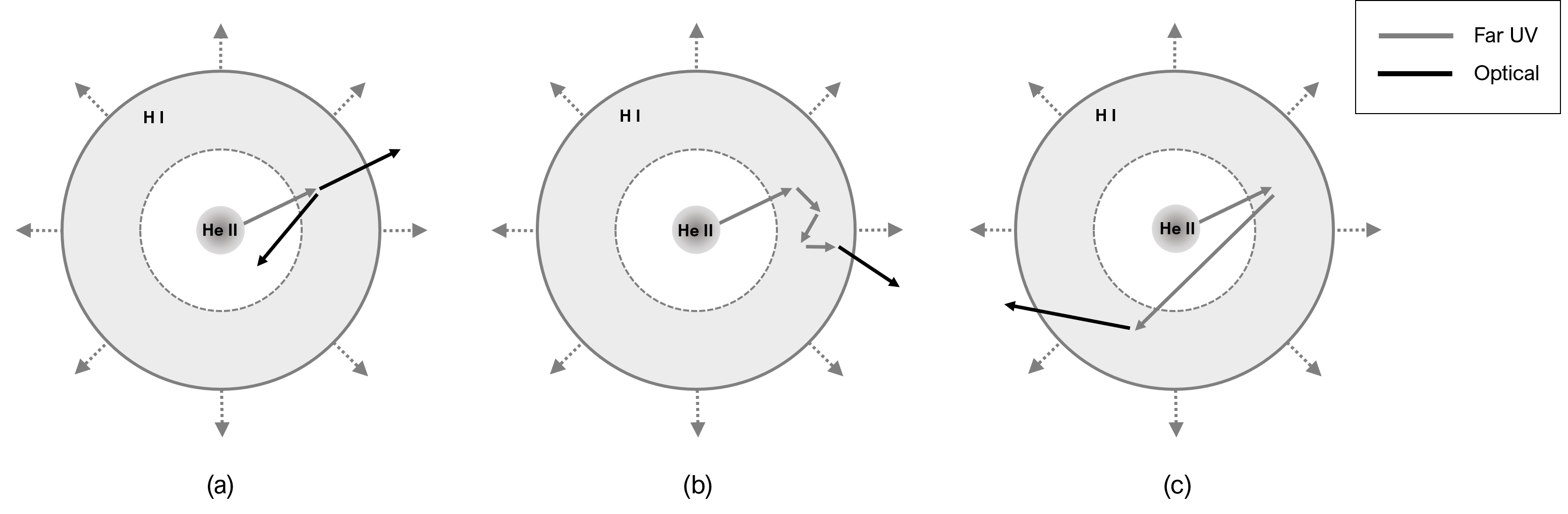}
\caption{Schematic illustration showing the formation of 
the double peak profile with an extended tail structure. The gray arrows 
indicate the path of far UV photons that are emitted from 
the central \ion{He}{2} source and subsequently Rayleigh scattered. 
The black arrows correspond to Raman-scattered optical \ion{He}{2} photons. 
%
(a) An escaping Raman photon is blueshifted for a forward final scattering
and a redshifted photon is obtained from a backward scattering.
(b) Local Rayleigh scattering results in only slight frequency diffusion.
(c) A Rayleigh reflection leads to significant wavelength increase leading
to formation of an extended red part.
}
\label{fig:scattering}
\end{figure}

In Fig.~\ref{fig:scattering}, we illustrate schematically how the main double 
peaks and the extended red part are formed. In panel~(a),
we show a singly scattered photon, where a forward moving Raman optical photon is detected by the observer as a photon constituting
the main blue peak with the Doppler factor $\Delta V_{\rm b}$ corresponding to $v_{\rm exp}$ from the Raman line center. 
On the other hand, a backward moving Raman photon 
falls on the main red peak with the wavelength corresponding to $-v_{\rm exp}$ from the Raman line center. In the optically thin limit,
due to symmetry between forward and backward scattering, we obtain a symmetric double peak profile.

Panel~(b) illustrates a case for an initially forward Rayleigh-scattered photon followed by a few local Rayleigh scatterings. In this case,
frequency diffusion is severely limited because of the local nature 
of scattering. This shows that in a highly optically thick case,
double peak profiles are restored. However, as shown in panel~(c), Rayleigh reflection at the inner surface may produce significant
redshift leading to formation of an extended red tail structure. 

\begin{figure}
\plottwo{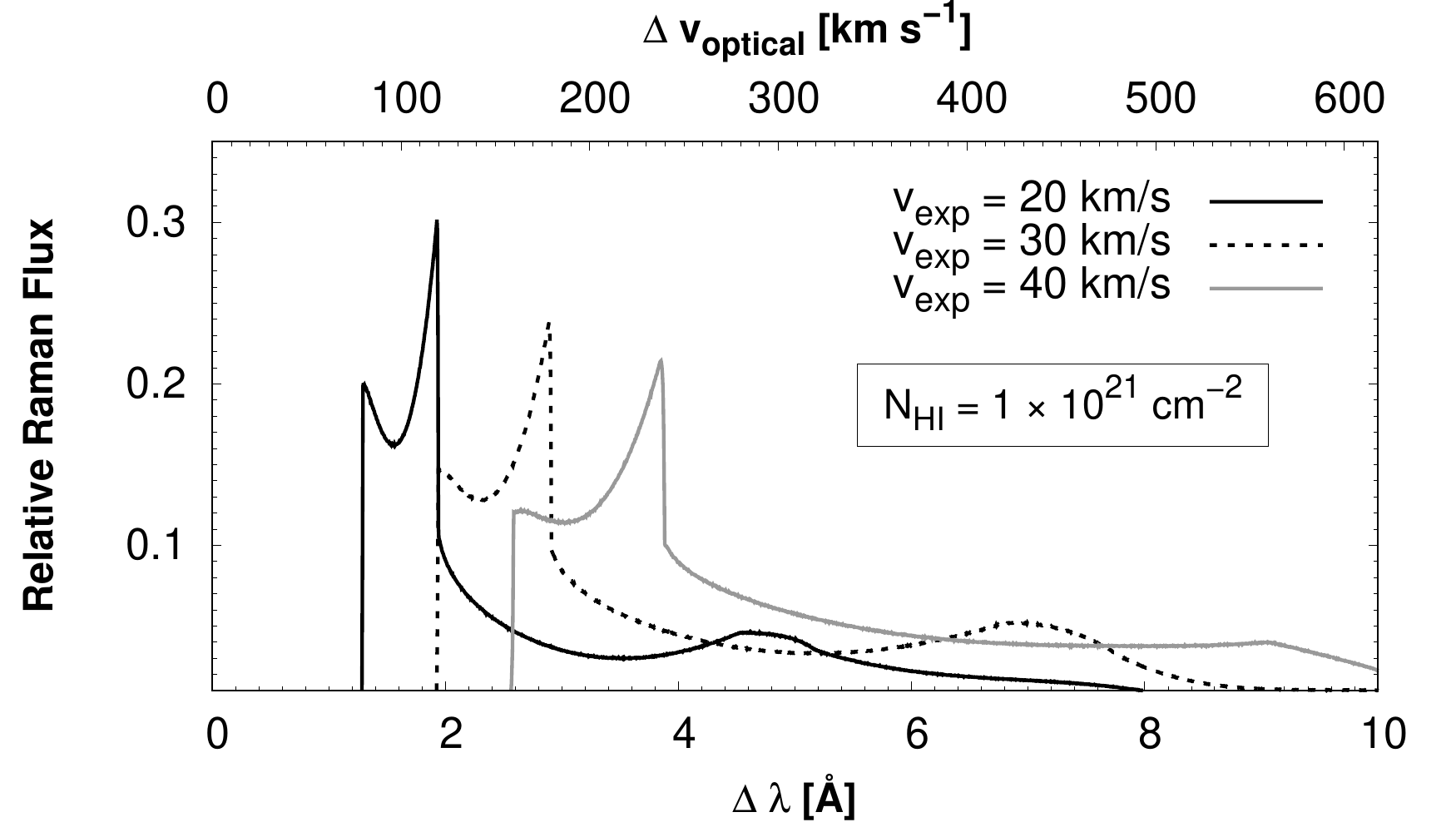}{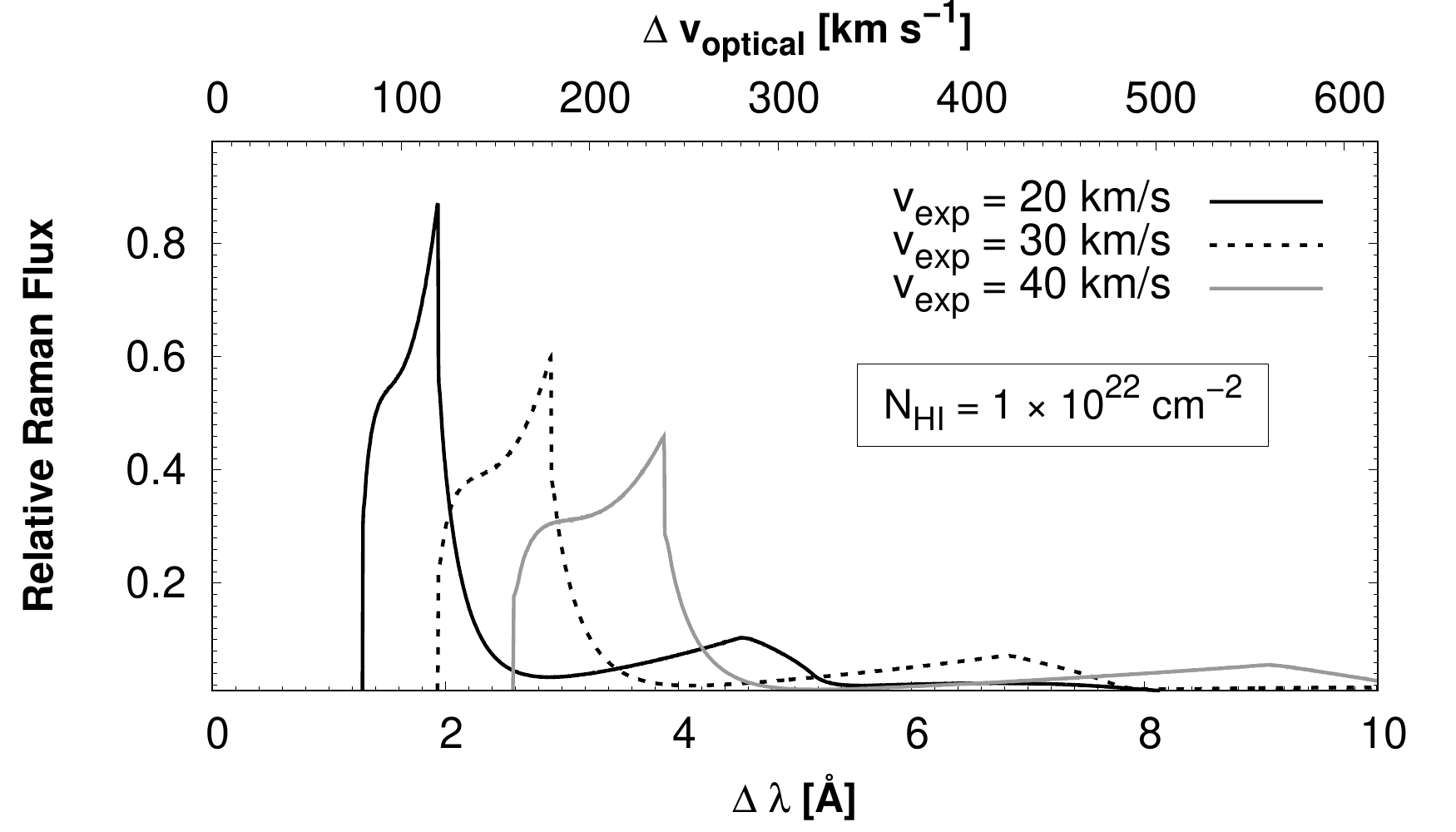}
\caption{
Line profiles of Raman \ion{He}{2} $\lambda$4851 obtained from our Monte Carlo simulations
for various values of the expansion speed $v_{\rm exp}$. The left and right panels show line profiles obtained for 
$N_{\rm HI} = 10^{21}{\rm\ cm^{-2}}$ and $10^{22}{\rm\ cm^{-2}}$. The horizontal and vertical axes are the same as
in Fig.~\ref{fig:mono_column}.
}
\label{fig:mono_vel}
\end{figure}

In Fig.~\ref{fig:mono_vel}, we present our simulated Raman \ion{He}{2} profiles for the values $20,30$ and $40{\rm\ km\ s^{-1}}$
of the expansion speed $v_{\rm exp}$
and for two fixed values $N_{\rm HI}=10^{21}{\rm\ cm^{-2}}$ and $10^{22}{\rm\ cm^{-2}}$ of \ion{H}{1} column density.
The resultant profiles exhibit a very asymmetric double peak structure with a tertiary peak in an extended red tail.
The center wavelength of Raman-scattered \ion{He}{2} shifts redward as $v_{\rm exp}$ increases. Furthermore, because
the cross section increases sharply toward \ion{H}{1} resonance, the line flux of Raman-scattered \ion{He}{2} also increases.
As we mentioned, the enhancement of red peaks and the formation of a tertiary peak are attributed to multiple Rayleigh scattering events 
before escape as an optical Raman photon. The tertiary peaks in the extended red tail region appear at $\Delta V=300, 450$
and $600{\rm\ km\ s^{-1}}$ for $v_{\rm exp}=20, 30$ and $40{\rm\ km\ s^{-1}}$, respectively. The tertiary peaks
are constituted by mostly twice Rayleigh-reflected at the inner surface at $r=R_i$ first at the part moving toward the observer
and secondly at the part moving away from the observer.

\begin{figure}
\plottwo{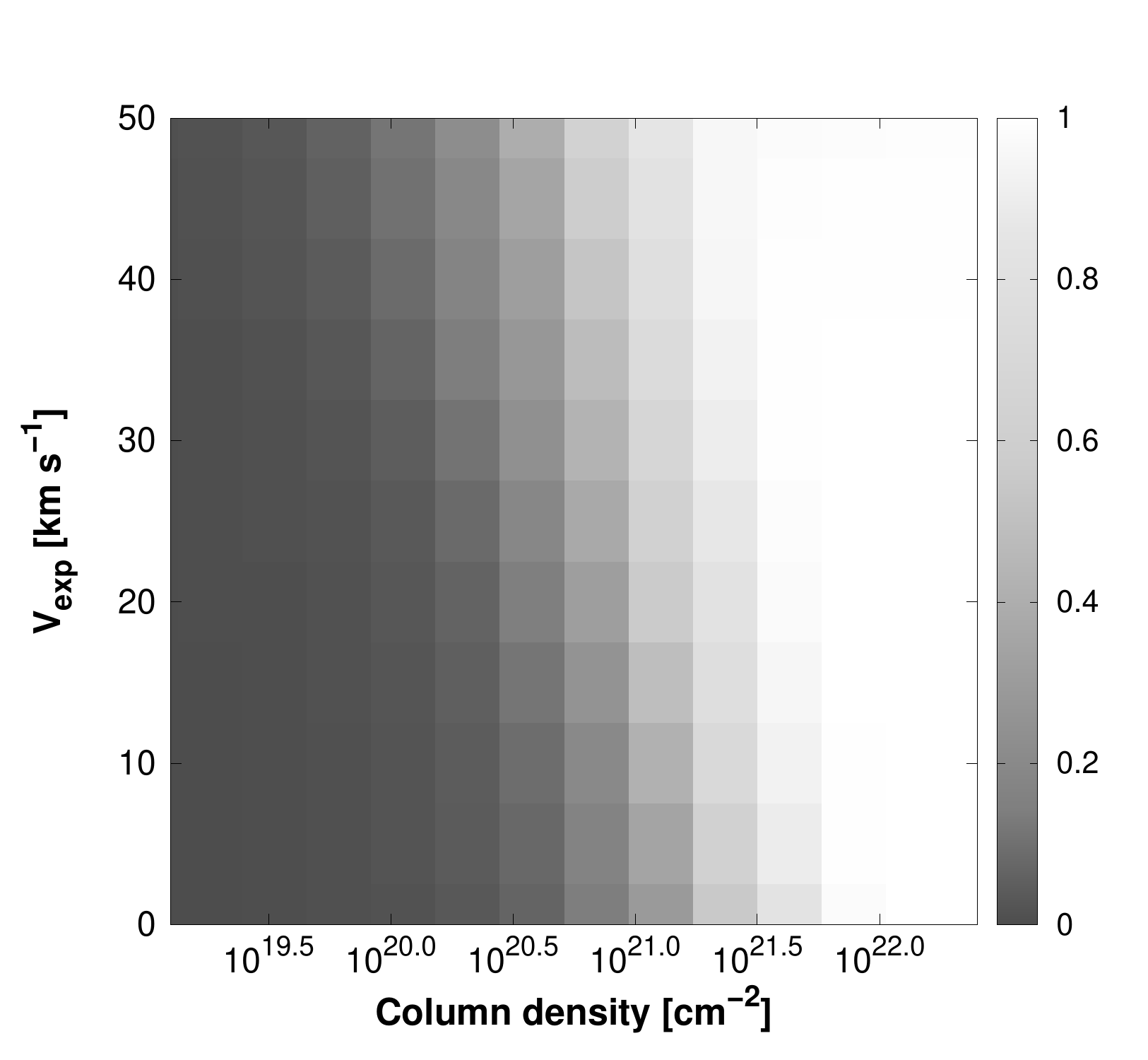}{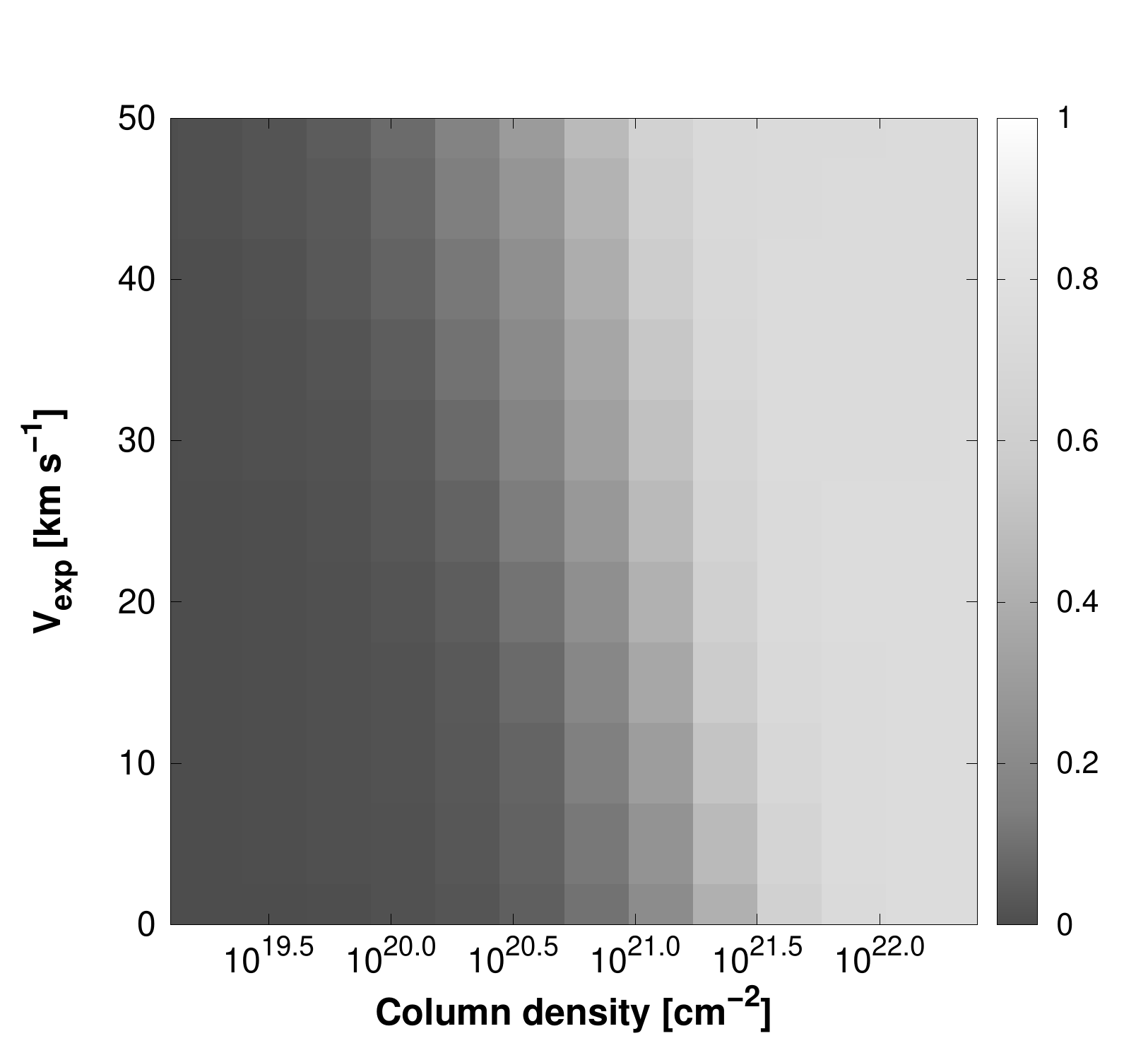}
\caption{Raman conversion efficiency for various values of $v_{\rm exp}$ and $N_{\rm HI}$.
The left panel shows the total Raman conversion efficiency $R_{\rm tot}$ for \ion{He}{2} $\lambda$972 defined as the sum of $R_{2s}$
and $R_{3s3d}$, the Raman conversion efficiencies into the $2s$ state and $3s, 3d$ states, respectively. The right panel shows the Raman conversion 
efficiency $R_{2s}$.
Note that the numerical results shown in the bottom of both panels correspond to
the case of a static \ion{H}{1} medium.
}
\label{fig:ram_conv}
\end{figure}

In Fig.~\ref{fig:ram_conv}, we plot the Raman conversion efficiency defined as the photon number ratio of incident 
\ion{He}{2}$\lambda$972 and Raman-scattered spectral line. For \ion{He}{2}$\lambda$972, there are two Raman scattering channels, 
one corresponding to final de-excitation into the $2s$ state and the other $3s$ and $3d$ states. Because in this work we only focus on line formation 
for Raman scattering into $2s$ state, we first compute the Raman conversion efficiency $R_{2s}$ defined as
\begin{equation}
\label{eq:R_2s}
R_{2s} = {\Phi_{2s} \over \Phi_{\rm HeII972}}
\end{equation}
where $\Phi_{\rm HeII972}$ and $\Phi_{2s}$ are number fluxes of Raman-scattered \ion{He}{2}$\lambda$4851 and far 
UV \ion{He}{2}$\lambda$972, respectively. In an analogous way, we define the Raman conversion efficiency $R_{3s3d}$ as
\begin{equation}
R_{3s3d} = {\Phi_{3s3d} \over \Phi_{\rm HeII972}},
\end{equation}
where $\Phi_{3s3d}$ is the number flux of Raman-scattered IR photons as a result of final de-excitation into $3s$ or $3d$ states.

In the left panel of Fig.~\ref{fig:ram_conv}, we show the total Raman conversion efficiency $R_{\rm tot}$ defined as the sum of 
$R_{2s}$ and $R_{3s3d}$. As is found in the left panel, the total Raman conversion efficiency $R_{\rm tot}$ increases to unity 
as either $N_{\rm HI}$ or $v_{\rm exp}$ increases. As $v_{\rm exp}$ increases, \ion{He}{2}$\lambda$972 line photons are 
redshifted toward Ly$\gamma$ in the rest frame of a hydrogen atom, resulting in dramatic increase of scattering cross section 
and sharp rise in Raman conversion efficiency. Raman scattering optical depth also increases by simply increasing $N_{\rm HI}$. 

Specifically, we obtain the Raman conversion efficiencies $R_{2s} = 0.16$ and 
$R_{3s3d}=0.06$ for the case of a stationary \ion{H}{1} region with \ion{H}{1} column 
density $N_{\rm HI}= 10^{21}\ \rm cm^{-2}$. This implies that 78 percent 
of far UV \ion{He}{2} escape through Rayleigh scattering or without interaction.
In contrast, in the case of an expanding neutral region with
$v_{\rm exp} =20\ \rm{km\ s^{-1}}$ and the same $N_{\rm HI}$, we obtain
$R_{2s}= 0.34$ and $R_{3s3d}=0.12$. The efficiencies increase to $R_{2s}= 0.51$
and $R_{3s3d}=0.18$ for an expanding neutral region with a speed of 
and $40\ \rm{km\ s^{-1}}$, respectively.

The right panel of Fig.~\ref{fig:ram_conv} shows the Raman conversion 
efficiency $R_{2s}$ giving 
the fraction of Raman-scattered
\ion{He}{2}$\lambda$4851 only. The highest value that $R_{2s}$ may reach 
in the optically thick limit is shown to be $\sim 0.8$,
which is found in the right-top portion of the figure. 
On the other hand, in the opposite limit of low optical depth, 
the fraction $R_{2s}$ is $\sim 0.1$, which is the quantum mechanical 
branching ratio for single scattering.

\subsection{Gaussian source} \label{sec:gauss}
With the assumption that the \ion{He}{2} emission source is isotropic and that the case B recombination is valid, 
we may safely infer the line profile of far UV \ion{He}{2}$\lambda$972 by investigating optical \ion{He}{2} emission lines 
such as \ion{He}{2}$\lambda$4686 and \ion{He}{2}$\lambda$4859 \citep{hummer87}. 
However, it should be noted that the validity of this assumption is questionable 
when a \ion{He}{2} emission region is in an orderly motion such as rotation 
in a specified plane that coincides with the binary orbital plane or perpendicular 
to the symmetry axis along which bipolar nebular morphology develops. 

\cite{jung04} performed a profile analysis of \ion{He}{2}$\lambda$4859 and H$\beta$ of the symbiotic star V1016~Cygni. 
They reported that H$\beta$ and \ion{He}{2}$\lambda$4859 are fitted well 
using a single Gaussian function with a full width at half maximum(FWHM) of $\Delta v_{\rm G}=77{\rm\ km\ s^{-1}}$. 
Considering the 4 times heavier atomic weight of \ion{He}{2} than hydrogen, 
the emission line profiles are attributed to convolution of thermal and turbulent motions. 
A similar study of the planetary nebula IC~5117 conducted by \cite{lee06} shows 
that H$\beta$ and \ion{He}{2}$\lambda$4859 are also well fitted by 
a single Gaussian function with FWHM of $\Delta v_{\rm G}=43{\rm\ km\ s^{-1}}$ and $35{\rm\ km\ s^{-1}}$, respectively. 
It appears that the profile widths of H$\beta$ and \ion{He}{2} are significantly contributed by
dynamical motions in addition to thermal motions.

\begin{figure}
\plottwo{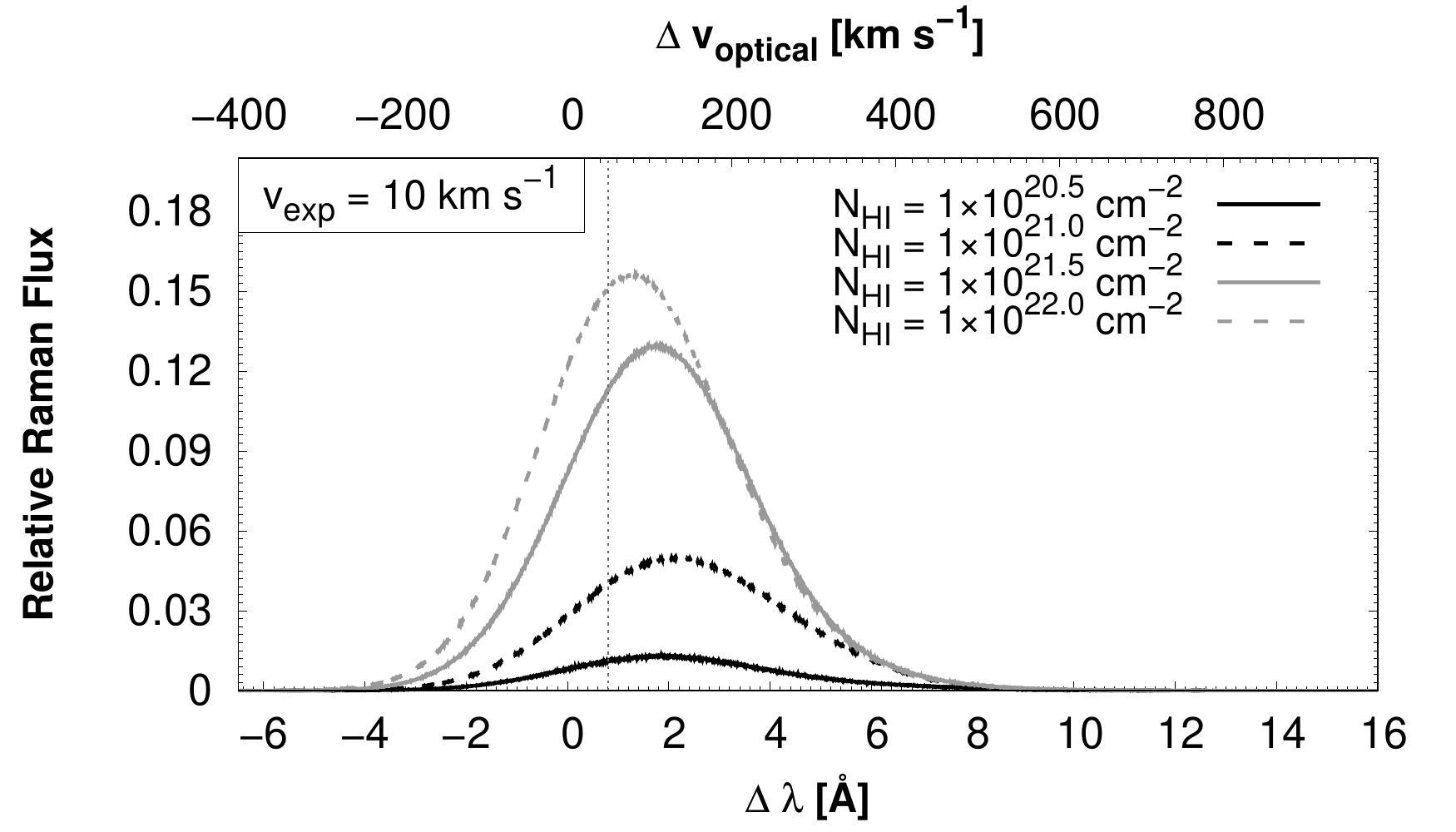}{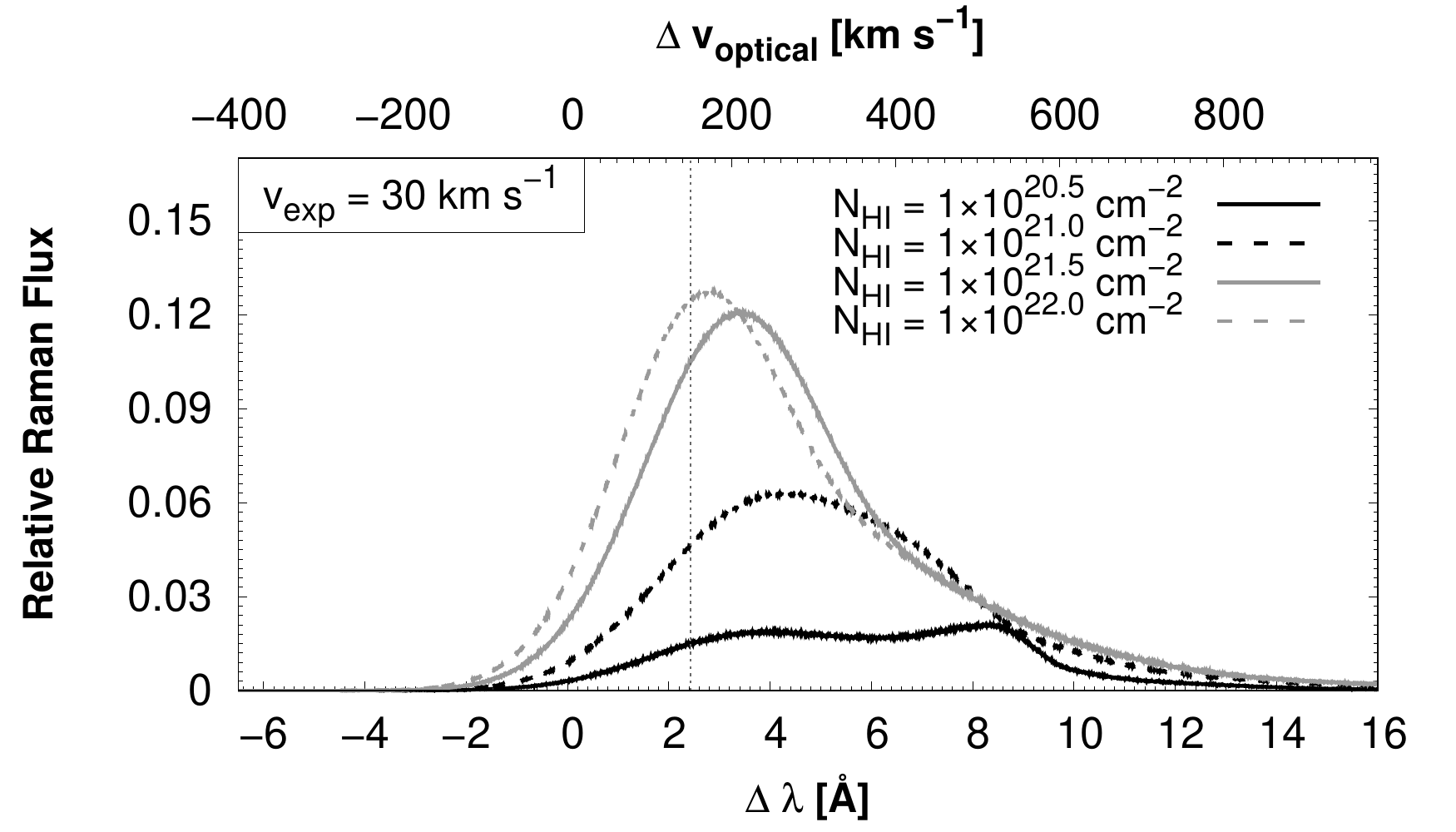}
\caption{Line profiles of Raman-scattered \ion{He}{2} formed in an expanding spherical neutral shell illuminated by a point-like
\ion{He}{2} line source with a Gaussian line profile with the FWHM of $\Delta v_{\rm G} = 50 {\rm\ km\ s^{-1}}$.
The left panel shows the resultant profiles for an expansion speed $v_{\rm exp}=10{\rm\ km\ s^{-1}}$ and 4 values of
$N_{\rm HI}$. The right panel is for $v_{\rm exp}=30{\rm\ km\ s^{-1}}$.
}
\label{fig:gauss_col}

\end{figure}

In Fig.~\ref{fig:gauss_col}, we show the line profiles of Raman-scattered \ion{He}{2}, 
where the \ion{He}{2} emission source is at the center with a Gaussian line profile with the FWHM 
$\Delta v_{\rm G}=50{\rm\ km\ s^{-1}}$. 
In the left panel, we show our result for an expansion speed $v_{\rm exp}=10{\rm\ km\ s^{-1}}$ and
the right panel shows the result for $v_{\rm exp}=30{\rm\ km \ s^{-1}}$. 
The dashed vertical line represents the wavelength that a line center 
\ion{He}{2} photon is Raman-scattered by a hydrogen atom 
moving away with $v_{\rm exp}$. The lower horizontal axis shows
the wavelength shift measured from $\lambda_{c,4851}$ defined in Eq.~\ref{eq:atomic_cen}.

In the left panel, we find that the line profiles are slightly
distorted redward because of sharp increase 
of Raman cross section. In the right panel, 
we find more severe profile distortion, which is attributed to 
the increased range of cross section variation. 
Multiply Rayleigh-scattered photons 
contribute to the emergent flux significantly in the red part 
in a complicated way.

In the case of $v_{\rm exp} = 10 {\rm\ km\ s^{-1}}$, 
the 'Raman line center' represented by the dashed vertical line 
is found at $\Delta\lambda = 0.81 \rm\ \AA$. The profile
peaks are formed at $\Delta \lambda = 1.83, 2.00, 1.79$, 
and $1.22 {\rm\ \AA}$ for $N_{\rm HI} = 10^{20.5}, 10^{21}, 10^{21.5}$, 
and $10^{22} \rm\ cm^{-2}$, respectively.
It should be noted that the peak position $\Delta\lambda$ varies 
with $N_{\rm HI}$ in a nonmonotonic way
\citep{jung04}.

On the other hand, the 'Raman line center' is found at 
$\Delta \lambda = 2.43 \rm\ \AA$ in the right panel, where $v_{\rm exp}=
30{\rm\ km\ s^{-1}}$.
We find that the peaks are located at
$\Delta \lambda = 4.23, 3.44$, and $2.83 {\rm\ \AA}$ for 
$N_{\rm HI} = 10^{21}, 10^{21.5}$, and $10^{22} \rm\ cm^{-2}$.
In the case of $N_{\rm HI}=10^{20.5}{\rm\ cm^{-2}}$, a conspicuous red peak 
at $\Delta \lambda = 8.36{\rm\ \AA}$ appears in addition to a peak at $\Delta\lambda= 4.12 {\rm\ \AA}$.

\begin{figure}
\plottwo{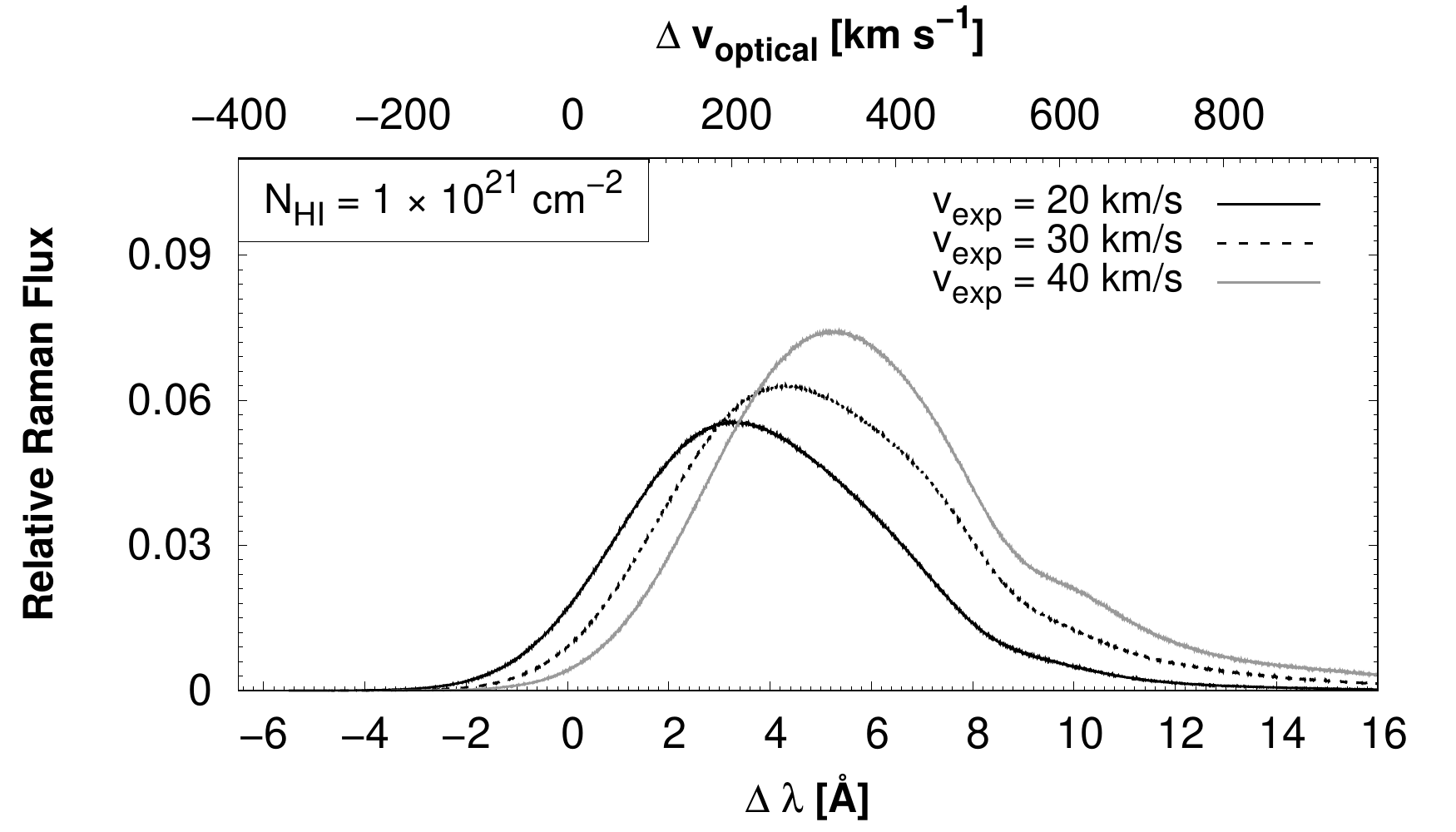}{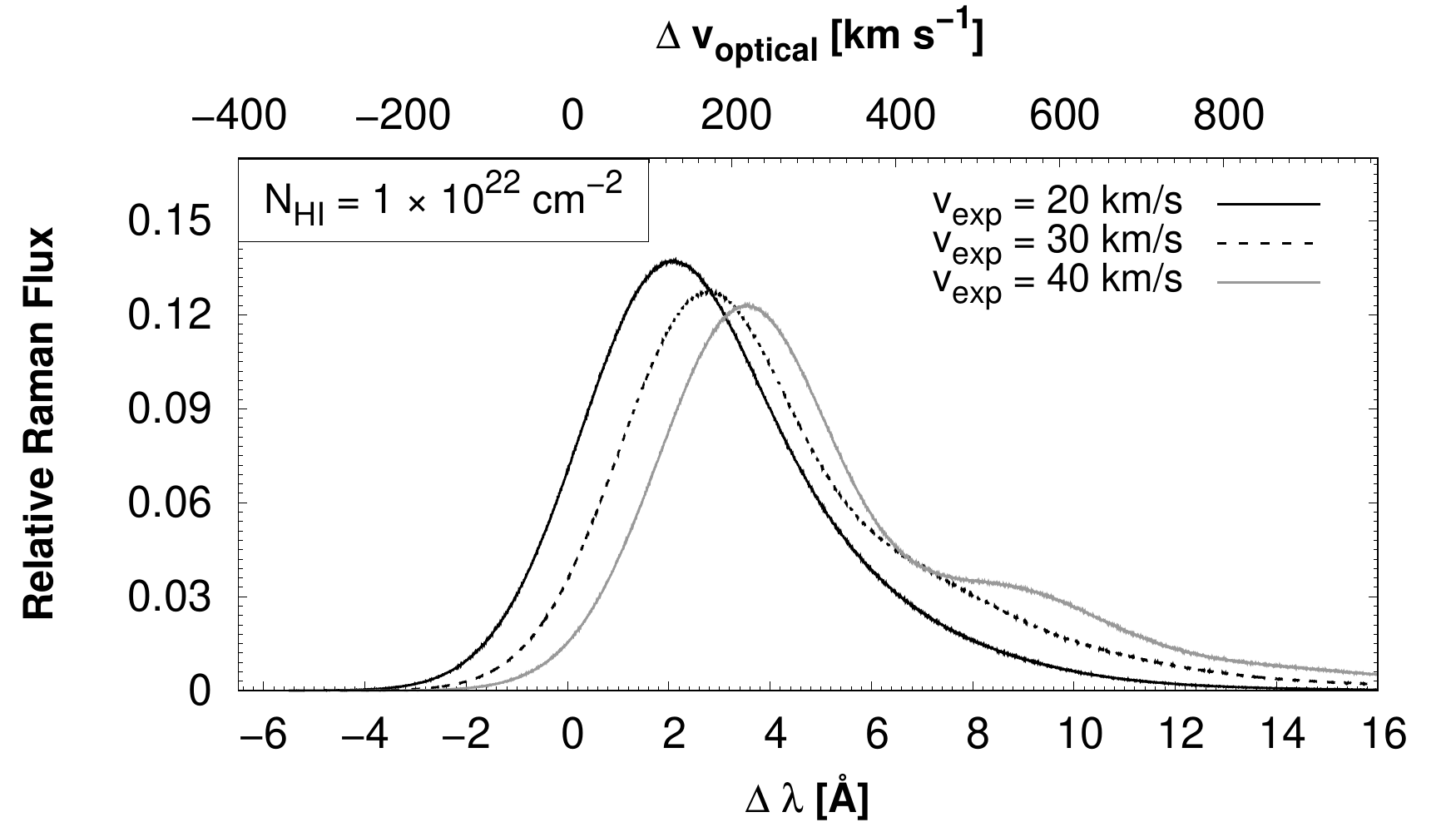}
\caption{The same result as Fig.~\ref{fig:gauss_col} but for various values
$v_{\rm exp}$.
The left panel shows the resultant profiles for $N_{\rm HI} = 10^{21} \rm cm^{-2}$ and 3 values of
$v_{\rm exp}$. The right panel is for $10^{22}{\rm\ cm\ s^{-2}}$. }
\label{fig:gauss_vel}
\end{figure}

In Fig.~\ref{fig:gauss_vel}, we present the line profile of Raman-scattered \ion{He}{2} with the same emission source
for a fixed value of $N_{\rm HI} = 10^{21} \rm\ cm^{-2}$
and $N_{\rm HI} = 10^{22} \rm\ cm^{-2}$ on the left and right panel, respectively. 
In the left panel, profile peaks are found at 
$\Delta \lambda = 3.23, 4.23$, and $5.29 {\rm\ \AA}$ 
for $v_{\rm exp} = 20, 30$, and $40 {\rm\ km\ s^{-1}}$. 
In the right panel with $N_{\rm HI}= 10^{22} \rm\ cm^{-2}$, we find that 
$\Delta \lambda = 2.04, 2.83$, and $3.52 {\rm\ \AA}$ 
for the same values of $v_{\rm exp}$ as in the left panel. 
We note that the peaks are not equally spaced and that
the line profiles are significantly skewed showing redward enhancement. 
This indicates that a great caution should be exercised in the determination of
the expansion speed from observed Raman-scattered \ion{He}{2} features.

\begin{figure}
\epsscale{1.2}
\plotone{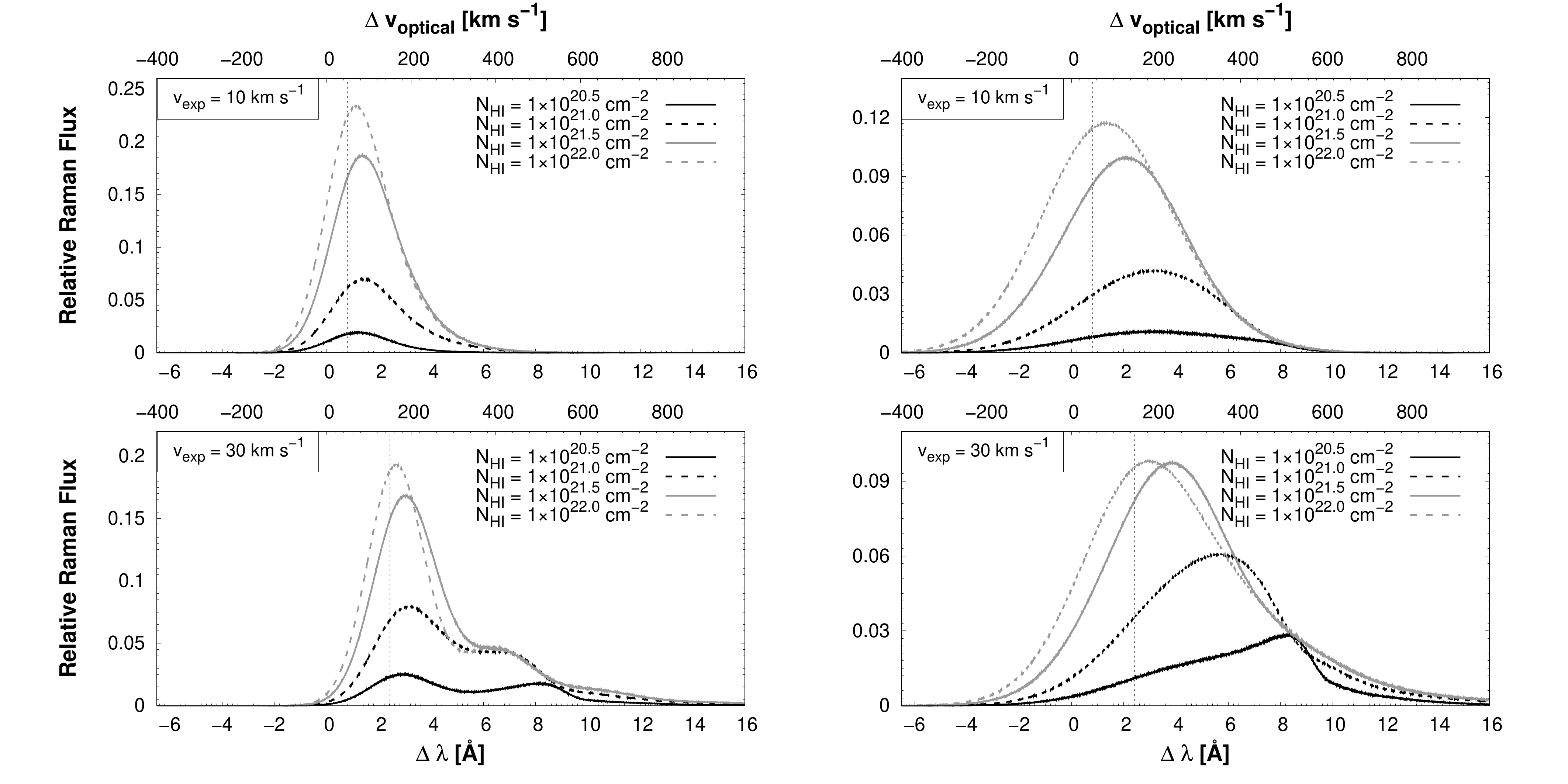}
\caption{
Line profiles of Raman-scattered \ion{He}{2} formed in an expanding spherical neutral shell 
with a point-like \ion{He}{2} source with a Gaussian line profile of which 
the FWHM $\Delta v_{\rm G} = 30{\rm\ km\ s^{-1}}$(left) and $70{\rm\ km\ s^{-1}}$(right).
The upper panel shows the resultant profiles for an expansion speed 
$v_{\rm exp} = 10{\rm\ km\ s^{-1}}$ with 4 values of $N_{\rm HI}$. The lower panel is for 
$v_{\rm exp} = 30{\rm\ km\ s^{-1}}$.
}
\label{fig:gauss_col_panel}
\end{figure}

\begin{figure}
\epsscale{1.2}
\plotone{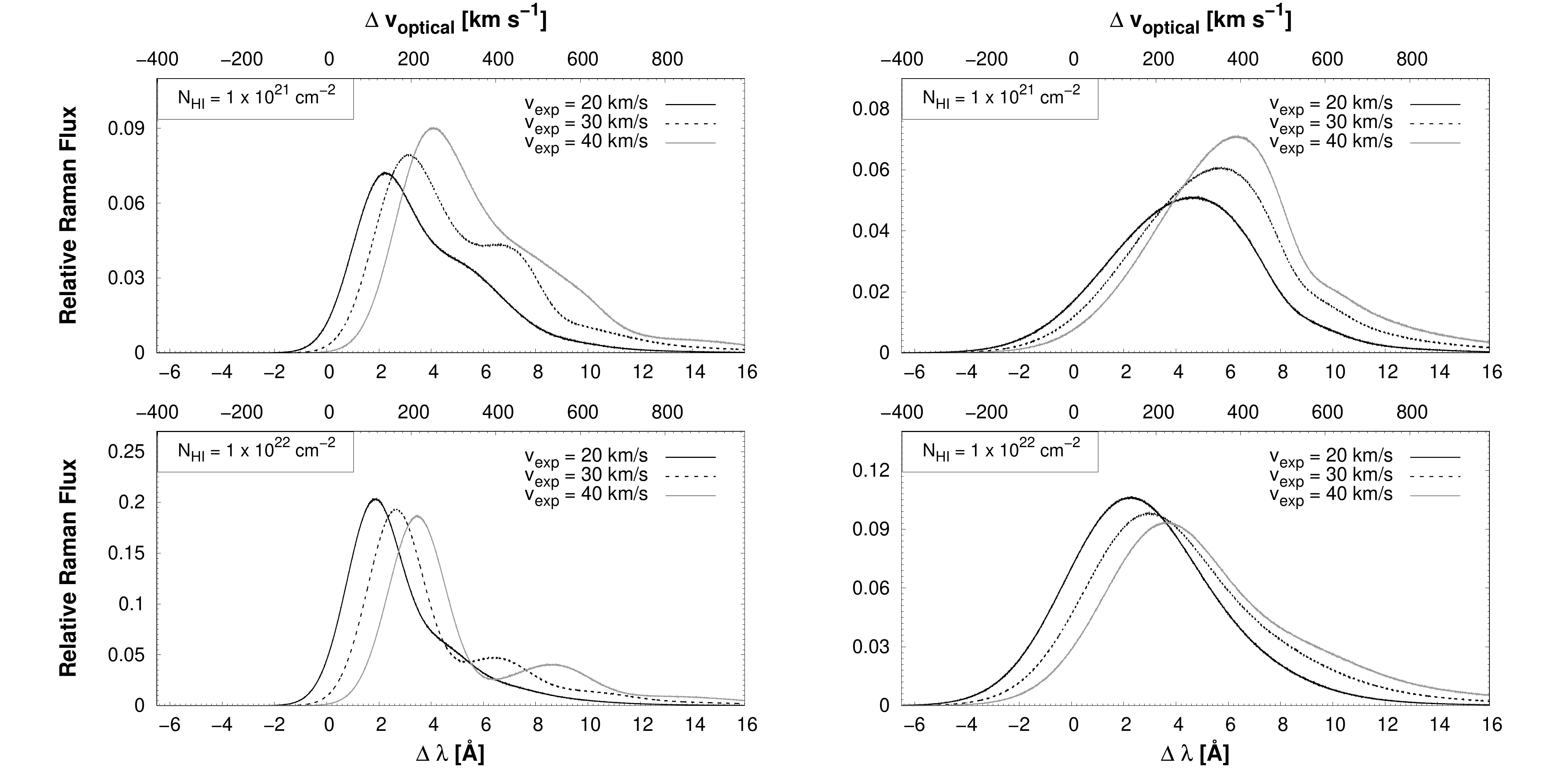}
\caption{
Line profiles of Raman-scattered \ion{He}{2} formed in an expanding spherical neutral shell 
with a point-like \ion{He}{2} source with a Gaussian line profile of which 
the FWHM $\Delta v_{\rm G} = 30{\rm\ km\ s^{-1}}$(left) and $70{\rm\ km\ s^{-1}}$(right).
The upper panel shows the resultant profiles for column density 
$N_{\rm HI} = 10^{21}{\rm\ cm^{-2}}$ with 3 values of expansion speed $v_{\rm exp}$. 
The lower panel is for $N_{\rm HI} = 10^{22}{\rm\ cm^{-2}}$.
}
\label{fig:gauss_vel_panel}
\end{figure}

We also investigate the dependence of $\Delta v_{\rm G}$ and show additional 
results for $\Delta v_{\rm G} = 30{\rm\ km\ s^{-1}}$ and
$70{\rm\ km\ s^{-1}}$ in Figs.~\ref{fig:gauss_col_panel} and \ref{fig:gauss_vel_panel}. 
We show the results for various values of $N_{\rm HI}$ in
Fig.~\ref{fig:gauss_col_panel}. In addition, Fig.~\ref{fig:gauss_vel_panel} shows the results
for various values of $v_{\rm exp}$. The left panels of Figs.~\ref{fig:gauss_col_panel} and
\ref{fig:gauss_vel_panel} are for $\Delta v_{\rm G}=30{\rm\ km\ s^{-1}}$
and the right panels show the results for 
$\Delta v_{\rm G}=70{\rm\ km\ s^{-1}}$.
In the top panels of Fig.~\ref{fig:gauss_col_panel} with $v_{\rm exp}=10{\rm\ km\ s^{-1}}$, we find
that all the line profiles are singly peaked with redward asymmetry. The nonmonotonic behavior
of the peak positions is similar to that found in the case of $\Delta v_{\rm G}=50{\rm\ km\ s^{-1}}$
as discussed in subsection \ref{sec:gauss}. For $\Delta v_{\rm G}=70{\rm\ km\ s^{-1}}$ and
$N_{\rm HI}=10^{21}{\rm\ cm^{-2}}$, the peak position is quite significantly redshifted and the line
profile is conspicuously distorted.

In the bottom left panel of Fig.~\ref{fig:gauss_col_panel} with $v_{\rm exp}=30{\rm\ km\ s^{-1}}$, additional red 
peaks are found. An incident \ion{He}{2} photon acquires a Doppler factor corresponding to $\sim 3 v_{\rm exp}$
as a result of double Rayleigh reflections in the spherical shell, which leads to the formation of the additional
red peaks. Because of the increased cross section for redshifted photons, the profile distortion is quite conspicuous for
$N_{\rm HI}=10^{20.5}{\rm\ cm^{-2}}$. 

In the bottom right panel of Fig.~\ref{fig:gauss_col_panel}, the resultant line profiles are singly peaked but
with large redward skewness. This result indicates that peak position may move redward substantially 
for broad \ion{He}{2} because line photons are distributed up to blue vicinity of Ly$\gamma$ resonance. 
In the case of $v_{\rm exp} = 30 \rm\ km\ s^{-1}$, 
$\Delta v_{\rm G} = 70~\rm km\ s^{-1}$ and $N_{\rm HI} = 10^{21} \rm\ cm^{-2}$, 
the profile peak is found at 
$\Delta\lambda = 5.54 \rm\ \AA$ that corresponds to $+343 \rm\ km\ s^{-1}$ in 
optical velocity space. It gives expansion speed of neutral medium about 
$v_{\exp} = 78 \rm\ km\ s^{-1}$ with a simple single Gaussian profile fitting. 
For $N_{\rm HI}\le 10^{21}{\rm\ cm^{-2}}$, the profile peaks are found at Doppler factors 
two or three times larger than that for $v_{\rm exp}=30{\rm\ km\ s^{-1}}$. This corroborates our conclusion
that $v_{\rm exp}$ should not be deduced from the peak position alone. 


In the top left panel of Fig.~\ref{fig:gauss_vel_panel} for which $N_{\rm HI}=10^{21}{\rm\ cm^{-2}}$, we observe development of red shoulder features
that are formed due to Rayleigh reflections at the inner surface of the \ion{H}{1} shell. As $v_{\rm exp}$ increases,
the peak positions move redward, which is accompanied with enhancement of Raman conversion efficiency.
This indicates very clearly that Raman conversion efficiency is affected by kinematics as well as the \ion{H}{1} column density
and covering factor. 

A similar behavior can be noticed in the top right panel, where the peak positions
appear at larger $\Delta\lambda$ than those in the left panel. It should be noticed that even the peak positions are
severely affected by the profile of the \ion{He}{2} emission source located at the center of the spherical \ion{H}{1} shell.

In the bottom panels of Fig.~\ref{fig:gauss_vel_panel} for which $N_{\rm HI}=10^{22}{\rm\ cm^{-2}}$, the development 
of an extended red tail part is severely suppressed compared to 
the case shown in the top panels. The line flux of Raman-scattered \ion{He}{2} 
is concentrated near the main peak
formed at $\Delta\lambda$ that corresponds to $v_{\rm exp}$. This is explained by the local nature of Rayleigh scattering that takes place near the entry spot 
in a scattering region with a large scattering optical depth. Only a small fraction of line photons make a substantial excursion by Rayleigh reflection
constituting the extended red part.

In the bottom left panel, Rayleigh reflected \ion{He}{2} photons form 
a relatively weak red peak at $(2-3)\times v_{\rm exp}$. 
However, in the right panels, the overall line profiles are singly peaked 
with a rather smooth extended tail in the red part. 
One may also notice that the overall Raman conversion
efficiency reaches the theoretical maximum value $\sim 0.8$ and quite independent of $v_{\rm exp}$. Due to suppression of development in the red part,
the peak positions faithfully represent the values of $v_{\rm exp}$.

\subsection{Application to the Young Planetary Nebula IC~5117} \label{subsec:observation}

\begin{figure}
\epsscale{1}
\plottwo{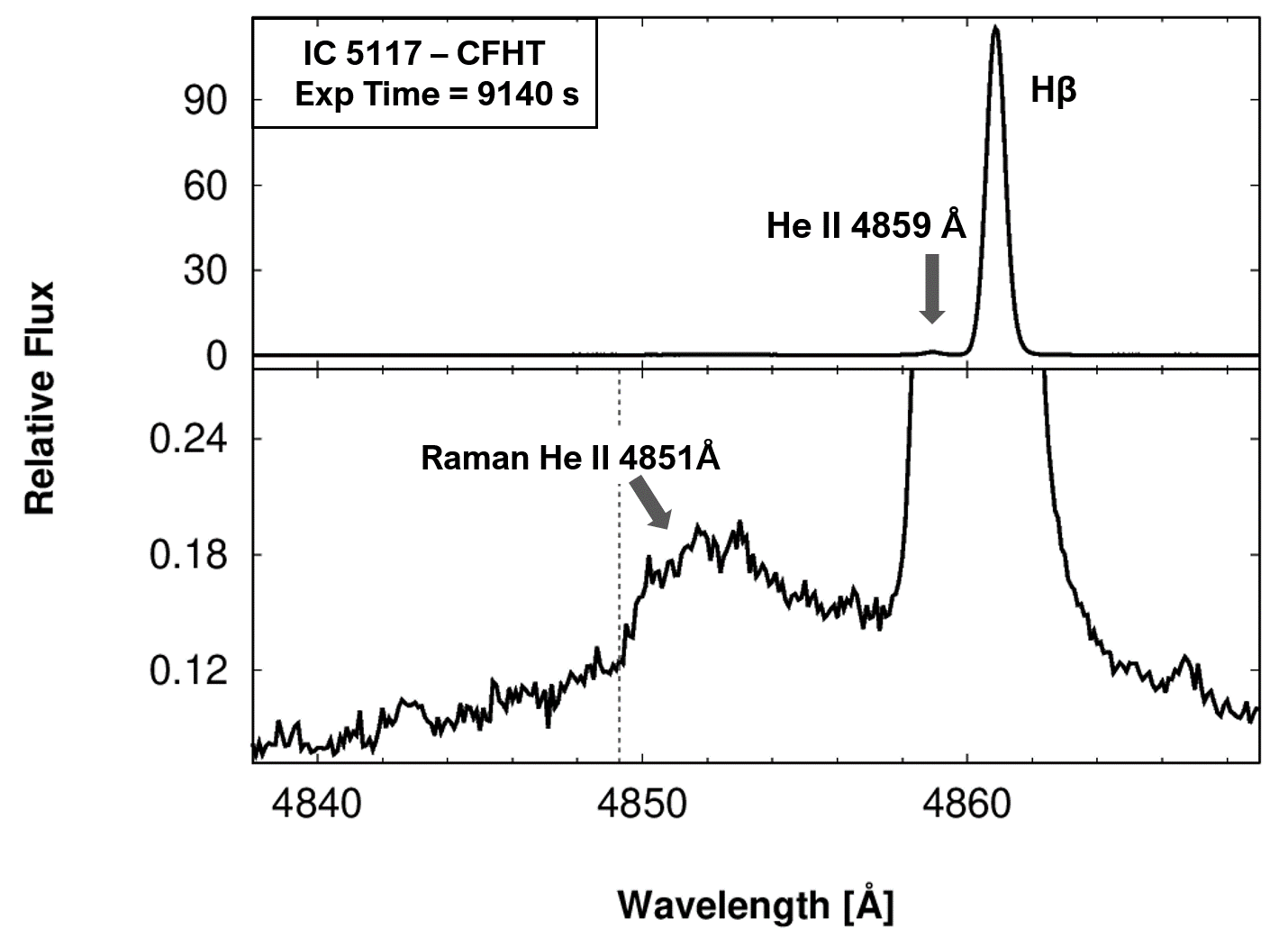}{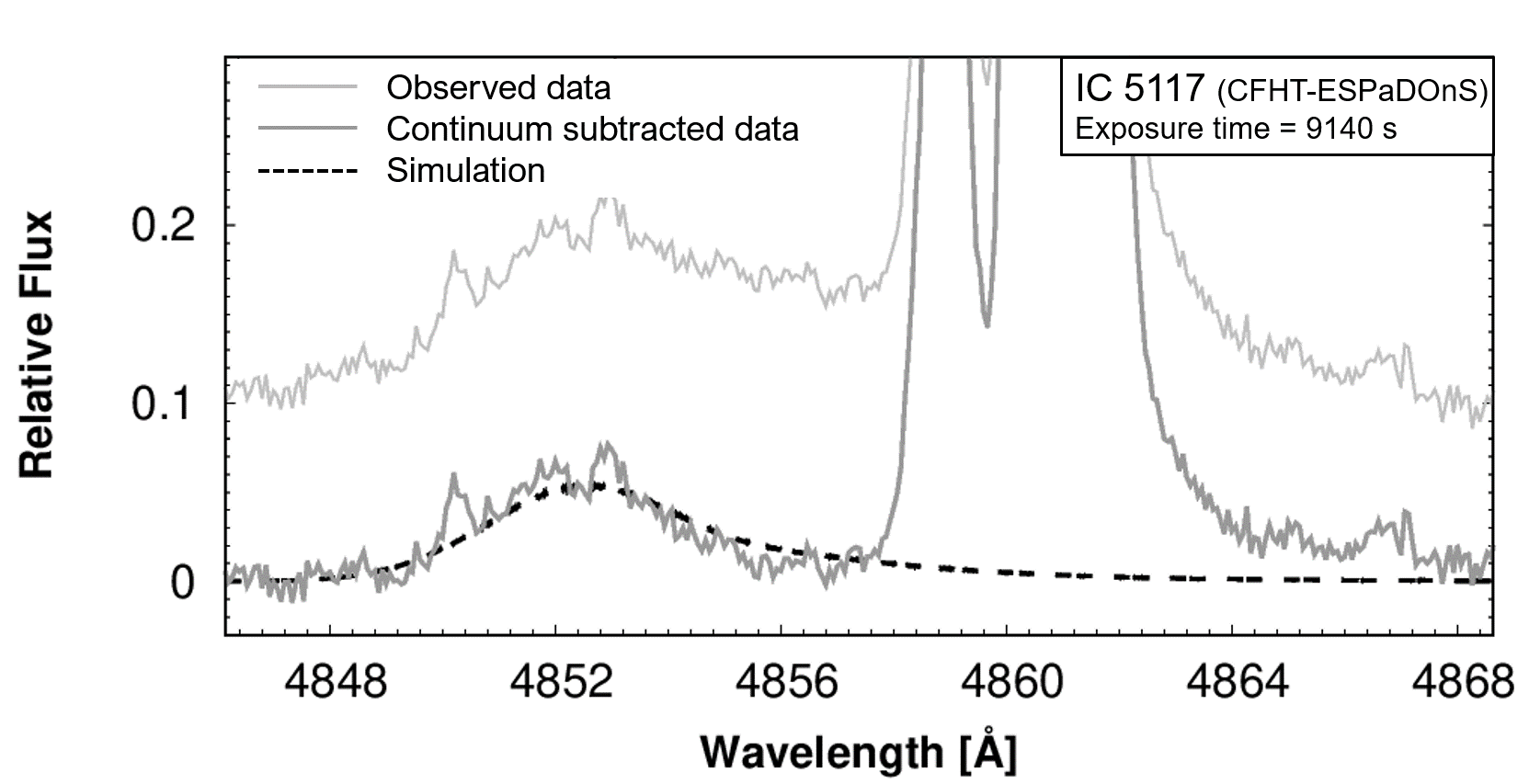}
\centering
\caption{{\it CFHT} spectrum of IC~5117 around H$\beta$ (left) and
our Monte Carlo profile fit (right).
\ion{He}{2}$\lambda$4859 is seen blueward of H$\beta$ in the upper left panel.
The lower panel is a blow-up version of the upper panel in order for clear
presentation of the very weak 
Raman-scattered \ion{He}{2} $\lambda$ 4851 appearing as a broad emission 
feature bluward of \ion{He}{2} $\lambda$ 4859. The right panel shows
continuum subtracted data and the best fit with simulated data obtained with 
$v_{\rm exp} = 30 \rm\ km\ s^{-1}$ and $\rm N_{\rm HI} = 10^{21} \rm\ cm^{-2}$.
}
\label{fig:ic5117}
\end{figure}

In the left panels of Fig~\ref{fig:ic5117}, we present a part of 
the spectrum showing Raman \ion{He}{2}$\lambda$4851 and H$\beta$
of IC~5117 observed with ESPaDOnS(Echelle Spectropolarimetric Device for the Observation of Stars)
installed on Canada-France-Hawaii~Telescope(CFHT). 
The observation was carried out on the two nights of September 6 and 14, 2014 
and the total exposure time is 9140~s.
The lower left panel is a blow-up version of the upper panel, in which
the Raman-scattered \ion{He}{2} feature appears clearly as a broad emission 
feature
blueward of \ion{He}{2}$\lambda$4859. 
The dashed vertical line marks the 'Raman line center' 
of Raman \ion{He}{2}$\lambda$4851,
which is expected to appear based on pure atomic physics relation given by
Eq~(\ref{eq:freq}).

The line profiles of H$\beta$, \ion{He}{2}$\lambda$4859 and
Raman-scattered \ion{He}{2}$\lambda$4851 are analyzed using a single Gaussian 
function 
\begin{equation}
f(\lambda)=f_0 \exp\left(-{{(\lambda - \lambda_0)^2} \over {2 \Delta \lambda}^2}\right),
\end{equation}
where the fitting parameters are presented in Table~\ref{tab:gaussian}. 
The line flux ratio of Raman \ion{He}{2} and \ion{He}{2} 
was 0.081 in the previous study of IC~5117 observed in 2004 by \cite{lee06},
which is different from the value 0.056 presented in Table~\ref{tab:gaussian}. 
However, this difference is partly due to the poor quality of 2004 data
and partly from ambiguity of continuum subtraction around weak and broad Raman \ion{He}{2}.

\begin{deluxetable*}{ccCrlc}[b!]
\tablecaption{
Single Gaussian parameters for observational data of IC~5117.
\label{tab:gaussian}}
\tablecolumns{6}
\tablewidth{0pt}
\tablehead{
\colhead{Line} &
\colhead{$\lambda_0$} &
\colhead{$\Delta \lambda$} & \colhead{$f_0$} \\
\colhead{} & \colhead{(\AA)} & \colhead{(\AA)} & \colhead{}
}
\startdata
H$\beta$~$\lambda$4861 & 4860.879 & 0.668 & 110.54 \\
\ion{He}{2}~$\lambda$4859 & 4858.942 & 0.573 & 1 \\
Raman~\ion{He}{2}~$\lambda$4851 & 4852.055 & 3.424 & 0.056 \\
\enddata
\end{deluxetable*}

The line center of Raman \ion{He}{2} is observed at 4852.06 \AA,
which is redward of the value expected from atomic physics by about 2.43 \AA.
This may indicate the expanding \ion{H}{1} region with a speed of 
$30 {\rm\ km\ s^{-1}}$.
The Raman conversion efficiency $R_{\rm 2s}$ is calculated with the same
method adopted by \cite{lee06} by assuming the validity of 
the case~B recombination theory for \ion{He}{2} 
\citep[e.g.][]{storey95}.
From the definition of $R_{\rm 2s}$ given in Eq~(\ref{eq:R_2s}) we have
\begin{equation}
R_{\rm 2s} \equiv {\Phi_{\rm 2s} \over \Phi_{\rm HeII972}}
= {{F_{\rm 4851} / h\nu_{\rm 4851}} \over {F_{\rm 972} / h\nu_{\rm 972}}}
= {\left({F_{\rm 4851} / h\nu_{\rm 4851}} 
\over {F_{\rm 4859} / h\nu_{\rm 4859}}\right)}
{\left({F_{\rm 4859} / h\nu_{\rm 4859}} 
\over {F_{\rm 972} / h\nu_{\rm 972}}\right)},
\end{equation}
where $F_{\rm 4851}$ is the total line flux of Raman \ion{He}{2}$\lambda$4851 
and other total \ion{He}{2} emission line fluxes are defined in a similar
way adopted by \cite{lee06}. The ratio of the photon number fluxes $\Phi_{\rm 4851} / 
\Phi_{\rm 4859}$ is measured to be 0.33.
Assuming the electron temperature $T_e = 2 \times 10^4~\rm K$ and the electron
density $n_e = 10^4~\rm cm^{-3}$, the case~B recombination theory yields the 
ratio of the photon number fluxes $\Phi_{\rm 4859} / \Phi_{\rm 972} = 1.046$, 
which finally leads to $R_{\rm 2s} = 0.34$.

In the right panel of Fig.~\ref{fig:ic5117}, we present our profile fit 
shown in dashed line. The adopted model parameters are 
$\Delta v_{\rm G} = 35{\rm\ km\ s^{-1}}$, 
$N_{\rm HI} = 10^{21} \rm cm^{-2}$ and $v_{\rm exp} = 30{\rm\ km\ s^{-1}}$. 
The overall agreement is acceptable and however, it should be noted
that the data is of insufficient quality.

This expansion speed is larger than the value $21{\rm\ km\ s^{-1}}$
suggested by \cite{weinberger89} who used [\ion{N}{2}] lines.
\cite{gussie95} used the Very Large Array to propose that 
the CO component is expanding with a speed $17{\rm\ km\ s^{-1}}$. 
The discrepancy in expansion speeds between our result and other researchers
may result from our adoption of a very simple
scattering geometry.   

However, one may notice that the Doppler factor associated with a Raman
feature is affected by the relative motion between the emitter and
the hydrogen atom and almost independent of the observer's line of sight.
If an emission region is aspherical such as toroidal and/or bipolar 
with an inclination angle $i$ with respect to the
observer's line of sight, direct emission spectroscopy yields somewhat
lower velocity width involving the factor $\sin i$. 
In contrast, atomic Raman spectroscopy reflects
only the relative speed of \ion{H}{1} and \ion{He}{2} with no underestimation
of the expansion speed due to the $\sin i$ factor. 

IC~5117 exhibits a bipolar nebular morphology and the
\ion{H}{1} region would be approximated by a cylindrical shell more
appropriately than a spherical one \citep{hsia14}. More reasonable
analyses can be performed with an adoption of a cylindrical shell model
including the projection effects to the celestial sphere.
We will investigate this model with consideration of polarization
in the near future.

\section{Summary and Discussion} \label{sec:discussion}

We investigate line formation of Raman-scattered \ion{He}{2}$\lambda$4851 in an 
expanding spherical \ion{H}{1} shell that may be found in young planetary nebulae, and symbiotic stars. 
In this work, we take into a careful consideration the change in cross section 
as the Doppler factor of a photon varies along the propagating path in a moving 
medium. In a spherically expanding medium, all the hydrogen atoms move away from each other. 
Therefore, in the rest frame of any hydrogen atom, line photons get redshifted 
toward Ly$\gamma$ resonance at which the scattering cross section increases sharply. 
This leads to significant enhancement of Raman conversion efficiency compared to
the case of a static \ion{H}{1} medium, as is illustrated in Fig.~\ref{fig:ram_conv}.
Another notable effect of expansion of the neutral region is frequency redistribution 
that tends to strengthen the red part 
of the emergent Raman \ion{He}{2} feature by systematically redshifting line photons.

It is found that the line profiles are mainly 
characterized by an asymmetric double peak structure with a significant
red tail that may extend to line centers of \ion{He}{2}$\lambda$4859 and H$\beta$. 
The extended tail part is contributed by photons that have acquired significant 
Doppler factor through one or a few Rayleigh-reflections at the inner surface of the neutral shell.
Blending with these two strong emission lines may severely hinder observational investigation of the extended red
tail part. 
In particular, young planetary nebulae and symbiotic stars exhibit broad
wings around Balmer emission lines, which also provides difficulty in isolating a clear profile of Raman-scattered \ion{He}{2}$\lambda$4851
\citep{chang18, lee00}.
Incomplete subtraction of Balmer wings may lead to an erroneous estimate of Raman conversion efficiency.

In this work, a new grid-based Monte Carlo code has been developed in order
to take into account the \ion{H}{1} density variation along a photon path.
This code is quite flexible so that it will be adopted to investigate
line formation of Raman-scattered \ion{O}{6} at 6825 \AA\ and 7082 \AA\
found in many symbiotic stars. In these objects, Raman scattering takes place
in a slow stellar wind from the giant component suffering a heavy mass loss.
Important kinematical information of the \ion{H}{1} component 
will be revealed through detailed line 
profile analyses of Raman \ion{O}{6} features 
\citep[e.g.,][]{schmid96, heo16, lee19}. 

Due to resonance nature in scattering 
cross section and branching ratios varying on wavelength, 
we obtain very diverse line profiles and Raman conversion efficiencies.
In this respect, it is very desirable to obtain high quality spectroscopic 
data of objects exhibiting Raman-scattered \ion{He}{2} features.
With these high quality data we may expect to secure faint Raman \ion{He}{2} features
formed blueward of high Balmer series lines including H$\gamma$ and H$\delta$. Much more
detailed information will be gained if one can compare as many Raman-scattered
\ion{He}{2} features as possible using instruments with wide spectral coverage.
For example, the line centers of Raman \ion{He}{2} are affected by both atomic
physics and kinematics of the \ion{H}{1} region. One way to isolate one effect from the other is 
to compare line centers of two or more Raman \ion{He}{2} features.


A significant fraction of planetary nebulae exhibit bipolar nebular morphology
\citep{sahai11}, for which circumnebular neutral region is approximated
by a cylindrical shell rather than a spherical shell. 
In the case of a neutral cylindrical shell, the emergent Raman-scattered \ion{He}{2}
features can be polarized dependent on the \ion{H}{1} column density and the orientation of the cylinder with respect to the observer's 
line of sight. In this case, the emergent profiles also depends on the observer's 
line of sight.

Considering that young planetary nebulae reported for detection of Raman-scattered \ion{He}{2} are also molecular line emitters,
it will be very interesting to carry out high resolution spectroscopy of molecular lines.
\cite{taylor90} suggested three possible mechanisms responsible for the formation 
of a neutral atomic region. One mechanism invokes atomic stellar winds that may occur in the AGB stage, while
the others involve the dissociation of molecular 
stellar winds either by UV radiation from the hot central star or by the ambient interstellar UV radiation field.
In particular, if the \ion{H}{1} region is inside the molecular region, the kinematical properties of the neutral shell
and molecular region are strongly correlated. Two dimensional spectroscopy using integrated field unit is expected to provide
interesting observational data.

\section*{Acknowledgements}
The authors are very grateful to the anonymous referee, who provided
constructive and helpful comments.
This work is based on observations obtained at the Canada-France-Hawaii 
Telescope (CFHT) which is operated by National Research Council of Canada, 
the Institut National des Sciences de l'Univers of the Centre National de 
la Recherche Scientifique of France, and the University of Hawaii.
This research was supported by the Korea Astronomy and Space Science Institute under the R\&D program(Project No. 2018-1-860-00) 
supervised by the Ministry of Science, ICT and Future Planning. This work was also supported by the National Research
Foundation of Korea(NRF) grant funded by the Korea government(MSIT) (No. NRF-2018R1D1A1B07043944).
This work was also supported by K-GMT Science Program 
(PID: cfht\_14BK002) funded through Korea GMT Project operated by Korea Astronomy and Space Science Institute.

\end{document}